\documentclass[twocolumn]{aastex631}
\usepackage{color}

\newcommand{\hii}{\mbox{\rm \ion{H}{2}}}
\newcommand{\co}{\mbox{\rm $^{12}$CO}}
\newcommand{\coone}{\mbox{\rm $^{12}$CO($1\text{--}0$)}}
\newcommand{\cotwo}{\mbox{\rm $^{12}$CO($2\text{--}1$)}}
\newcommand{\tco}{\mbox{\rm $^{13}$CO}}
\newcommand{\tcoone}{\mbox{\rm $^{13}$CO($1\text{--}0$)}}
\newcommand{\tcotwo}{\mbox{\rm $^{13}$CO($2\text{--}1$)}}
\newcommand{\ceo}{\mbox{\rm C$^{18}$O}}
\newcommand{\ceoone}{\mbox{\rm C$^{18}$O($1\text{--}0$)}}
\newcommand{\ceotwo}{\mbox{\rm C$^{18}$O($2\text{--}1$)}}
\newcommand{\aCO}{\mbox{$\alpha_{\rm CO}$}}
\newcommand{\aLTE}{\mbox{$\alpha_{\rm LTE}$}}
\newcommand{\aISM}{\mbox{$\alpha_{\rm ISM}$}}
\newcommand{\Tkin}{\mbox{$T_{\rm kin}$}}
\newcommand{\nhydro}{\mbox{$n_{\rm H_2}$}}
\newcommand{\Nhydro}{\mbox{$N_{\rm H_2}$}}
\newcommand{\Ssfr}{\mbox{$\Sigma_{\rm SFR}$}}
\newcommand{\Sssc}{\mbox{$\Sigma_{\rm SSC}$}}
\newcommand{\Shydro}{\mbox{$\Sigma_{\rm H_2}$}}
\newcommand{\uvolume}{\mbox{cm$^{-3}$}}
\newcommand{\kms}{\mbox{km s$^{-1}$}}
\newcommand{\Kkms}{\mbox{K km s$^{-1}$}}
\newcommand{\Jyb}{\mbox{Jy beam$^{-1}$}}
\newcommand{\Jykms}{\mbox{Jy km s$^{-1}$}}
\newcommand{\Jybkms}{\mbox{Jy beam$^{-1}$ km s$^{-1}$}}

\newcommand{\uaco}{\mbox{$M_{\odot}$ (K km s$^{-1}$ pc$^2$)$^{-1}$}}
\newcommand{\usfr}{\mbox{$M_{\odot}$ yr$^{-1}$}}
\newcommand{\udsfr}{\mbox{$M_{\odot}$ kpc$^{-2}$ yr$^{-1}$}}
\newcommand{\udhydro}{\mbox{$M_{\odot}$ pc$^{-2}$}}
\newcommand{\Mihara}{\affil{Tokyo Metropolitan Mihara High School, 1-33-1 Omorihigashi, Ota-ku, Tokyo, Japan}}
\newcommand{\JWU}{\affil{Japan Women's University, 2-8-1 Mejirodai, Bunkyo-ku, Tokyo 112-8681, Japan}}
\newcommand{\MPIA}{\affil{Max-Planck-Institut f\"{u}r Astronomie, K\"{o}nigstuhl 17, D-69117, Heidelberg, Germany}}
\newcommand{\Nichidai}{\affil{Department of Physics, General Studies, College of Engineering, Nihon University, 1 Nakagawara, Tokusada, Tamuramachi, Koriyama, Fukushima 963-8642, Japan}}
\newcommand{\NAOJ}{\affil{National Astronomical Observatory of Japan, 2-21-1 Osawa, Mitaka, Tokyo 181-8588, Japan}}
\newcommand{\GUAS}{\affil{The Graduate University for Advanced Studies (SOKENDAI), 2-21-1 Osawa, Mitaka, Tokyo 181-0015, Japan}}
\newcommand{\MPE}{\affil{Max-Planck-Institut f\"{u}r Extraterrestrische Physik, Giessenbachstrasse, D-85748, Garching, Germany}}
\newcommand{\IoA}{\affil{Institute of Astronomy, The University of Tokyo, 2-21-1 Osawa, Mitaka, Tokyo 181-0015, Japan}}
\newcommand{\Hokudai}{\affil{Department of Physics, Faculty of Science, Hokkaido University, Kita 10 Nishi 8 Kita-ku, Sapporo, Hokkaido 060-0810, Japan}}
\newcommand{\SAAO}{\affil{South African Astronomical Observatory, PO Box 9, Observatory 7935, Cape Town, South Africa}}
\newcommand{\UGranada}{\affil{Departamento de F\'{i}sica Te\'{o}rica y del Cosmos, Campus de Fuentenueva, Universidad de Granada, E-18071 Granada, Spain}}
\newcommand{\Joetsu}{\affil{Graduate School of Education, Joetsu University of Education, 1, Yamayashiki-machi, Joetsu, Niigata 943-8512, Japan}}
\newcommand{\OsakaU}{\affil{Department of Earth and Space Science, Osaka University, 1-1 Machikaneyama, Toyonaka, Osaka 560-0043, Japan}}
\newcommand{\UAntananarivo}{\affil{Department of Physics, University of Antananarivo, P.O. Box 906, Antananarivo, Madagascar}}

\received{July 2, 2021}
\revised{February 17, 2022}
\shorttitle{ALMA Observations of the Early-stage Interacting Galaxy NGC~3110}
\shortauthors{Y. Kawana et al.}

\graphicspath{{./}{figures/}}

\begin{document}
\title{Multi-wavelength and Multi-CO View of The Minor Merger Driven Star Formation in the Nearby LIRG NGC~3110}

\correspondingauthor{Toshiki Saito}
\email{toshiki.saito@nao.ac.jp, saito.toshiki@nihon-u.ac.jp}

\author{Yuka Kawana}\Mihara\JWU
\author[0000-0002-2501-9328]{Toshiki Saito}\Nichidai\NAOJ\MPIA
\author{Sachiko K. Okumura}\JWU
\author[0000-0002-8049-7525]{Ryohei Kawabe}\NAOJ
\author[0000-0002-8726-7685]{Daniel Espada}\UGranada
\author[0000-0002-2364-0823]{Daisuke Iono}\NAOJ\GUAS
\author[0000-0002-2699-4862]{Hiroyuki Kaneko}\Joetsu\NAOJ
\author[0000-0002-2419-3068]{Minju M. Lee}\MPE
\author[0000-0003-2475-7983]{Tomonari Michiyama}\OsakaU\NAOJ
\author[0000-0002-0724-9146]{Kentaro Motohara}\IoA\NAOJ
\author[0000-0002-0724-9146]{Kouichiro Nakanishi}\NAOJ\GUAS
\author[0000-0002-3662-3942]{Alex R. Pettitt}\Hokudai
\author[0000-0003-2666-4158]{Zara Randriamanakoto}\SAAO\UAntananarivo
\author[0000-0003-3652-495X]{Junko Ueda}\NAOJ
\author[0000-0002-4999-9965]{Takuji Yamashita}\NAOJ

%%%%%%%%%%%%%%%%%%%%%%%%%%%%%%
%%%%%%%%%% Abstract %%%%%%%%%%
%%%%%%%%%%%%%%%%%%%%%%%%%%%%%%
\begin{abstract}
We present Atacama Large Millimeter/submillimeter Array observations of multiple \co, \tco, and \ceo\ lines and 2.9~mm and 1.3~mm continuum emission toward the nearby interacting luminous infrared galaxy NGC~3110, supplemented with similar spatial resolution H$\alpha$, 1.4~GHz continuum, and $K$-band data. We estimate the typical CO-to-H$_2$ conversion factor of 1.7~\uaco\ within the disk using LTE-based and dust-based H$_2$ column densities, and measure the 1-kpc scale surface densities of star formation rate (\Ssfr), super star clusters (\Sssc), molecular gas mass, and star formation efficiency (SFE) toward the entire gas disk. These parameters show a peak at the southern part of the southern spiral arm (SFE $\sim$ 10$^{-8.2}$~yr$^{-1}$, $\Ssfr$ $\sim$ 10$^{-0.6}$\udsfr, $\Sssc$ $\sim$ 6.0~kpc$^{-2}$), which is likely attributed to the on-going tidal interaction with the companion galaxy MCG-01-26-013, as well as toward the circumnuclear region. We also find that thermal free-free emission contributes to a significant fraction of the millimeter continuum emission at the southern peak position. Those measurements imply that the peak of the southern arm is an active and young star-forming region, whereas the central part of NGC~3110 is a site of long-continued star formation. We suggest that, during the early stage of the galaxy-galaxy interaction with large mass ratio that in NGC~3110, fragmentation along the main galaxy's arms is an important driver of merger-induced star formation and massive gas inflow results in dusty nuclear starbursts.
\end{abstract}

%% https://astrothesaurus.org

%%%%%%%%%%%%%%%%%%%%%%%%%%%%%%
%%%%%%%% Introduction %%%%%%%%
%%%%%%%%%%%%%%%%%%%%%%%%%%%%%%
\section{Introduction} \label{sec:intro}
Studying star formation processes in various galaxies provides us critical information to unveil how galaxies are formed and evolved in time and thus to investigate the origin of the Hubble sequence \citep{Hubble26,Kennicutt98a}.
When two galaxies approach each other, they exert tidal forces and change their morphology. In the case of gas-rich galaxies, large-scale inflow supplies molecular gas to the central regions, resulting in the burst of nuclear star formation \citep[e.g.,][]{Barnes92,Mihos96}. On the other hand, recent observational and theoretical works have revealed that, at the early-to-mid stage of the interaction, the merger-driven star formation is more dominated by gas fragmentation across the progenitor's disks \citep[e.g.,][]{Teyssier10,Elmegreen17,Pettitt17,Tomicic18}, or in a filamentary structure between the progenitors \citep[e.g.,][]{Saitoh09,Iono13,Saito15,Kaneko18}.

In order to improve our understanding of the non-linear response of gas during a collision, we require high quality millimeter/submillimeter molecular line data which can trace the distribution and kinematics of cold molecular gas, the reservoir for future star formation. For nearby galaxies, low-$J$ \co, \tco, and \ceo\ lines ($n$ $\simeq$ 10$^2$-10$^3$~\uvolume) are usually employed to trace the total column density and the response of H$_2$ gas to violent merger events. The optically-thin \tco\ and \ceo\ lines are rather important to constrain physical properties of the diffuse molecular ISM.

\begin{figure}
\begin{center}
\includegraphics[width=8.5cm]{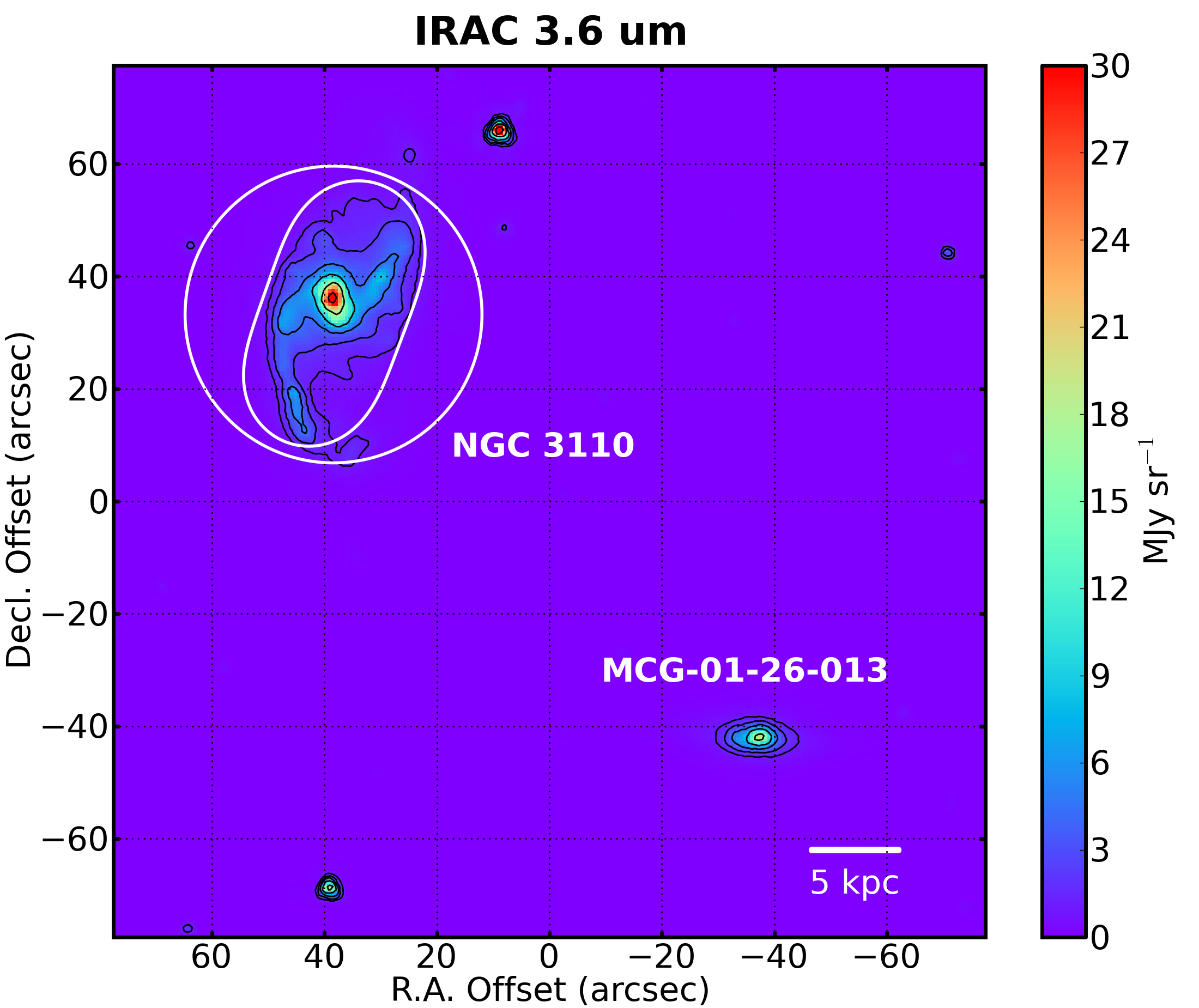}
\end{center}
\caption{The IRAC 3.6$\mu$m color image of NGC~3110. The circle and ellipse show the typical field of view of Band 3 and Band 6 observations, respectively. A companion galaxy MCG-01-26-013 can be seen in the lower right of NGC~3110.
}\label{fig:irac}
\end{figure}

A gas-rich nearby interacting galaxy NGC~3110 ($D_{\rm L}$ = 69.4~Mpc; 1\arcsec = 325~pc), which is classified as a nearby luminous infrared galaxy (LIRG) ($L_{\rm IR}$ = 10$^{11.1}$~$L_{\odot}$; \citealt{Armus09}), is one of the best targets to study starbursts in early-stage interactions (Figure~\ref{fig:irac}), because of its uniqueness as well as rich multi-wavelength observations and detailed simulations reproducing its interacting nature in the literature (e.g., \citealt{Espada18}). It consists of a central bar-like north-south elongation and a pair of asymmetric arms extending from the center, which are considered to be due to the tidal interaction with a companion galaxy MCG-01-26-013 located $\sim$38 kpc southwest from the center of NGC~3110 \citep{Espada18}. A recent near-IR survey \citep{Randriamanakoto13,Randriamanakoto15,Randriamanakoto17} reported the detection of more than 280 super star cluster (SSC) candidates across the disk, implying that NGC~3110 is one of the SSC-richest LIRGs in the low-redshift Universe. These SSCs, gravitationaly bound objects, are young and massive star clusters likely to form in strong starbursts and during galaxy-galaxy merging process \citep[e.g.,][]{Whitmore99,Randriamanakoto19}. Given its uniqueness and active star formation, studying NGC~3110 can provide valuable information for understanding how starbursts are triggered in early-stage mergers.

Another unique feature of this galaxy is the bright H$\alpha$ blob at the southern edge of the southern tidal arm \citep{Hattori04}. \citet{Espada18} mapped \cotwo\ with $\sim$1-kpc spatial resolution, and found an asymmetric gas distribution with a strong concentration around the nucleus, which coincide with the H$\alpha$ structures. They found that the average star formation efficiency (SFE) in the spiral arms is $\sim$0.5dex higher than that in the circumnuclear region, implying that off-centered starbursts are triggered in early stages of the merger. They also carried out hydrodynamical minor merger simulations in order to reproduce the morphological characteristics (e.g., rotation curve) seen in their \cotwo\ map and H$\alpha$ map. They concluded that the prominent asymmetric two-armed structures in NGC~3110 is formed at $\sim$150~Myr after the closest approach, and the system is experiencing one of the highest star formation rate (SFR) episodes as a result of the fly-by interaction with the minor companion galaxy.

Most of previous extragalactic molecular gas studies including NGC~3110 assumed a single \co\ luminosity to H$_2$ gas mass conversion factor (\aCO), which is likely to spatially vary among and within galaxies \citep{Bolatto13,Sandstrom13}, to derive the H$_2$ gas mass surface density and some relevant quantities such as SFE. In order to better understand the missing properties of molecular gas, including \aCO, and the relation to star formation in merging galaxies, we have conducted multi-CO ALMA observations toward NGC~3110.

This paper is organized as follows: the ALMA observations, the data reduction, and the photometry method are summarized in Section~\ref{sec:data} and results are briefly summarized in Section~\ref{sec:results}. We describe how we derive the physical properties, e.g., H$_2$ gas mass, kinetic temperature, extinction-corrected SFR, and SSC number density, in Section~\ref{sec:derive}. We discuss the star formation properties of NGC~3110 and their relation to the interaction event using the derived quantities (Section~\ref{sec:discussion}), and then summarize and conclude this Paper in Section~\ref{sec:summary}. We have adopted H$_0$ = 70 \kms\ Mpc$^{-1}$, $\Omega_m$ = 0.3, and $\Omega_{\Delta}$ = 0.7 throughout this Paper.

%%%%%%%%%%%%%%%%%%%%%%%%%%%%%%
%%%%%%%%%%%% Data %%%%%%%%%%%%
%%%%%%%%%%%%%%%%%%%%%%%%%%%%%%
\section{Observations and Data Reduction} \label{sec:data}
\subsection{ALMA Observations}
Observations of \co\ and its isotopologues toward NGC~3110 were carried out during the ALMA cycle 2 period (ID: 2013.1.01172.S). We set four spectral setups whose upper sideband were tuned to \co\ or \tco. The main targeted lines are $J$ = 1--0 (Band~3) and 2--1 (Band~6) transitions of \co, \tco, and \ceo. All data were obtained during 2014--2015. The single-sideband system temperature at Band~3 and Band~6 are 40--155~K and 65--135~K, respectively. The assigned antenna configurations for the Band~3 and Band~6 observations had 29--41 and 35--38 12~m antennas with the projected baseline length between 15--558~m and 15--348~m, respectively. Each tuning had four spectral windows (spws) to cover both sidebands. All spws had a bandwidth of 1.875~GHz with 7.812~MHz resolution. We used 3 pointings to cover all gaseous structures found in the \cotwo\ image taken by the Submillimeter Array \citep{Espada10,Espada18}, although one pointing is enough for the Band~3 observations. All observation setups are summarized in Table~\ref{table_obs}.

\subsection{Data Reduction}
Data calibration and imaging were done using the Common Astronomy Software Applications package ({\tt CASA}) version 4.2.2 \citep{McMullin07}. When we reconstructed a \cotwo\ image cube using the visibility data delivered from the observatory, we saw periodical linear patterns along one direction which were different from thermal noise structure. Thus, we carefully checked the visibility data and then flagged some data points whose amplitude deviated from the average value around a given baseline length. This additional flagging, as a result, suppressed the periodic patterns. We adopt this flagging reduction scheme to all ALMA data sets shown in this paper.

Images were reconstructed with the natural $uv$ (robustness parameter = 2.0) weighting and the velocity resolution of 20 or 40~\kms\ depending on the achieved signal-to-noise ratio of the targeted lines. Continuum emission was subtracted in the $uv$ domain by fitting the line-free channels in both upper sideband and lower sideband with a first order polynomial function. The line-free channels were used to make a continuum image using the multi-frequency synthesis method. We used the CASA task {\tt tclean} in multi-scale mode to make use of the multiscale CLEAN deconvolution algorithm \citep{Cornwell08}. All imaging properties for detected molecular lines including sensitivity and beam size are listed in Table~\ref{table_line}. Throughout this paper, we adopt the typical systematic errors on the absolute flux calibration of 5\% and 10\% for the Band~3 and Band~6 data, respectively \citep{Lundgren13}.

Before making line intensity ratio maps, we flagged the $uv$ range below 10~k$\lambda$, so that the maximum recoverable scale (MRS) of each ALMA image is the same. The MRS definition we used in this paper is,
\begin{equation}
{\rm MRS} \approx 0.6\frac{\lambda_{\rm obs}}{L_{\rm min}},
\end{equation}
where $\lambda_{\rm obs}$ is the observed wavelength and $L_{\rm min}$ is the minimum projected baseline length \citep{Lundgren13}. The minimum $uv$ distance is determined by the configuration of the \cotwo\ observations. Since the truncated $uv$ range at 10~k$\lambda$ corresponds to the MRS of $\sim$10\farcs7 (3.5~kpc), we do not consider the missing flux effect on structures smaller than 10\farcs7. However, we caution that our \coone\ data miss some extended emission (recovered flux $=$ 77 $\pm$ 15~\%) compared to single dish measurement (395 $\pm$ 79~\Jykms; \citealt{Sanders91}), which may affect our measurements on 3\farcs0 apertures, although fainter line data tend to be less affected. The aperture size is smaller than the MRS (described in Section~\ref{sec:photometry}). We made all maps and flux measurements after correcting for the primary beam attenuation. Then, we convolve all the ALMA line and continuum maps to 2\farcs0 resolution. We note that we use the original (i.e., untruncated) data for the gas mass measurement in this paper, although for all works requiring any line ratio we use the truncated data.

After imaging, we created moment maps following the steps described here: (1) clip the \coone\ datacube at 3$\sigma$ (see Table~\ref{table_line}), (2) convert the clipped cube to a 1/0 mask, (3) expand the mask by convolving by 2\farcs0 beam, and (4) collapse the datacubes after applying the expanded mask with 2.5$\sigma$ clipping for the $^{12}$CO data and 1.5$\sigma$ for the other fainter data. This procedure allows us to include faint structures neighboring strong structures, as well as exclude patchy faint structures which are likely artifacts.

\begin{figure*}[t]
\begin{center}
\includegraphics[width=18cm]{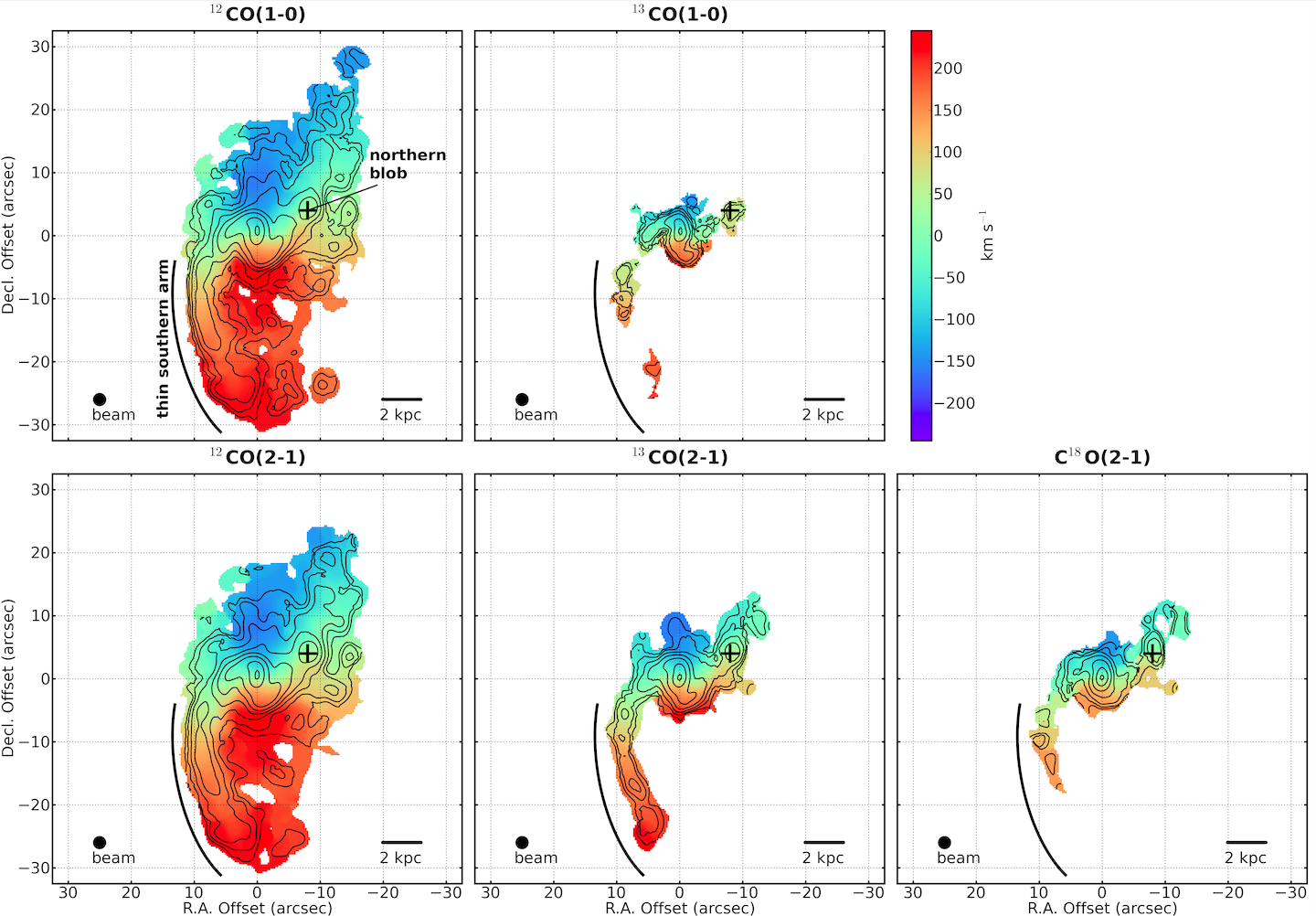}
\end{center}
\caption{(top-left) \coone\ integrated intensity contours overlaid on \coone\ intensity-weighted velocity field color image. The contours are 21.6 $\times$ (0.01, 0.02, 0.04, 0.08, 0.16, 0.32, 0.64, and 0.96) \Jybkms. The center position of this image is ($\alpha$, $\delta$)$_{\rm J2000}$ = (10h04m02.090s, -6d28m29.604s).
(top-middle) \cotwo\ image. The contours are 67.1 $\times$ (0.01, 0.02, 0.04, 0.08, 0.16, 0.32, 0.64, and 0.96) \Jybkms.
(top-right) \tcoone\ image. The contours are 1.85 $\times$ (0.02, 0.04, 0.08, 0.16, 0.32, 0.64, and 0.96) \Jybkms.
(bottom-left) \tcotwo\ image. The contours are 5.70 $\times$ (0.02, 0.04, 0.08, 0.16, 0.32, 0.64, and 0.96) \Jybkms.
(bottom-right) \ceotwo\ image. The contours are 2.16 $\times$ (0.02, 0.04, 0.08, 0.16, 0.32, 0.64, and 0.96) \Jybkms.
}\label{fig:showline}
\end{figure*}

\subsection{Ancillary Data}
In addition to the ALMA data, we use the H$\alpha$ map obtained with Okayama Astrophysical Observatory (OAO) 188~cm telescope. The observations and data reduction are described in detail in \citet{Hattori04}. The reported spatial resolution is 1\farcs35, which is similar to the ALMA resolution. We also download the 1.4~GHz radio continuum map taken by Karl G. Jansky Very Large Array (VLA) from VLA Data Archive\footnote{\url{https://archive.nrao.edu/archive}}. These maps are used to construct an extinction-corrected SFR map (see Section~\ref{sec:derive}). Furthermore, we use a high-quality $K$-band map obtained by Very Large Telescope (VLT) (see \citealt{Randriamanakoto13} for details of the data), which provides the spatial distribution of SSCs.

\subsection{Nyquist-sampled Aperture Photometry}\label{sec:photometry}
In order to decrease the number of correlated data points and to unify the flux measurement method among images taken with different telescopes, a hexagonal Nyquist sampling (aperture diameter $=$ 3\farcs0, aperture separation $=$ 1\farcs5) was performed on all data taken by ALMA and complementary data from the OAO (H$\alpha$), VLT ($K$-band), and VLA (radio continuum) observations. In addition, the hexagonal gridding is one of the popular ways to analyze spatially-resolved extragalactic systems \citep[e.g.,][]{Sandstrom13,denBrok21}. The Nyquist-sampled data points with flux stronger than 3$\sigma$ will be used to derive physical parameters (Section~\ref{sec:derive}) and to reconstruct physical parameter images.

We converted the observed integrated line flux density, $S\Delta v$ (\Jybkms) to integrated line intensity, $I$ (\Kkms) using this conversion factor, $\beta$,
\begin{equation}
\frac{\beta}{\rm Jy\:beam^{-1}\:K^{-1}} = 1.222 \times 10^6\left(\frac{\theta_{\rm maj}\theta_{\rm min}}{\rm arcsec^2}\right)^{-2}\left(\frac{\nu_{\rm obs}}{\rm GHz}\right)^{-2},
\end{equation}
where $\theta_{\rm maj}$ and $\theta_{\rm min}$ are the FWHM sizes in the major and minor axis of the synthesized beam, respectively, and $\nu_{\rm obs}$ is the observed frequency.

The integrated line intensity per aperture, $I_{\rm area}$, was calculated by,
\begin{equation}
\frac{I_{\rm area}}{\rm K\:km\:s^{-1}} = \frac{\sum_{i}^{N_{\rm px,area}}I_{\rm px,i}}{N_{\rm px,area}},
\end{equation}
where $I_{\rm px,i}$ is the integrated line intensity at $i$-th pixel (\Kkms), $N_{\rm px,area}$ is the number of pixels within the 3\farcs0 aperture. $I_{\rm px,i}$ is same as $\beta(S\Delta v)_{\rm px,i}$, where $(S\Delta v)_{\rm px,i}$ is the observed line flux density at $i$-th pixel in \Jybkms\ unit. When calculating the ALMA and VLA continuum brightness temperatures per apertures, one just needs to replace $I_{\rm area}$, $I_{\rm px,i}$, $(S\Delta v)_{\rm px,i}$ with $T_{\rm b,area}$ (K), $T_{\rm b,px,i}$ (K), $(S_{\nu_{\rm obs}})_{\rm px,i}$ (\Jyb), respectively.

The statistical noise (i.e., sensitivity) per aperture, $\sigma_{\rm stat}(I_{\rm area})$, was calculated by the following equation \citep[e.g.,][]{Hainline04},
\begin{equation}
\frac{\sigma_{\rm stat}(I_{\rm area})}{\rm K\:km\:s^{-1}} = \beta\sigma_{\rm ch}V_{\rm ch}\sqrt{N_{\rm ch}},
\end{equation}
where $\sigma_{\rm ch}$ is the median single-channel rms per pixel (\Jyb), $V_{\rm ch}$ is the velocity width of a channel, and $N_{\rm ch}$ is the number of channels integrated over.
For the continuum data, we used,
\begin{equation}
\frac{\sigma_{\rm stat}(T_{\rm b,area})}{\rm K} = \beta\sigma_{\rm rms},
\end{equation}
where $\sigma_{\rm rms}$ is the median rms per pixel.

When we consider the systematic error of the absolute flux calibration, $\sigma_{\rm sys}$, as well as $\sigma_{\rm stat}$, we combined both errors using the equation below,
\begin{equation}
\sigma_{\rm tot} = \sqrt{\sigma_{\rm stat}^2 + \sigma_{\rm sys}^2},
\end{equation}
where we assume that both errors are independent. Note that all the maps shown in this paper are clipped based on $\sigma_{\rm stat}$, whereas the error bars in all the plots are calculated based on $\sigma_{\rm tot}$.

Hereafter, integrated flux density and integrated intensity of a certain line, for example \coone, will be written as $S_{\rm CO(1-0)}\Delta v$ and $I_{\rm CO(1-0)}$, respectively.

The line luminosity per aperture can be calculated from,
\begin{equation}
\frac{L'_{\rm line}}{\rm K\:km\:s^{-1}\:pc^2} = 3.25 \times 10^7S_{\rm line}\Delta v\:\nu_{\rm obs}^{-2}D_{\rm L}^{2}(1 + z)^{-3},
\end{equation}
where $D_{\rm L}$ is the luminosity distance ($=$ 69.4~Mpc), and $z$ is the redshift \citep{Solomon05}. In this paper, the \coone\ luminosity is used to calculate \aCO.

\begin{figure*}
\begin{center}
\includegraphics[width=15cm]{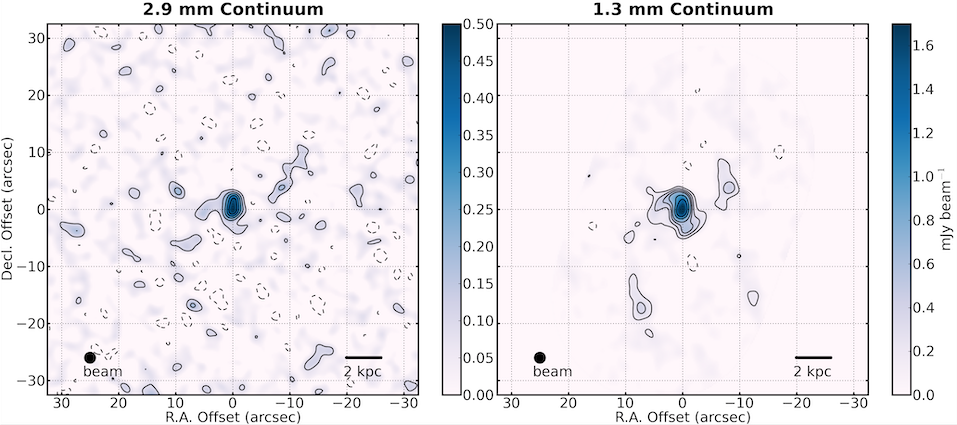}
\end{center}
\caption{(top) The 2.9~mm continuum image of NGC~3110. The contours are -2.0, 2.0, 4.0, 6.0, 8.0, 10.0$\sigma$. 1$\sigma$ corresponds to 41.8$\mu$Jy beam$^{-1}$. The center position of this image is the same as Figure~\ref{fig:showline}.
(bottom) The 1.3~mm continuum image. The contours are -2.5, 2.5, 5.0, 7.5, 10.0, 15.0, and 20.0$\sigma$. 1$\sigma$ corresponds to 62.7$\mu$Jy beam$^{-1}$.
 }\label{fig_contin}
\end{figure*}

%%%%%%%%%%%%%%%%%%%%%%%%%%%%%%
%%%%%%%%%% Results %%%%%%%%%%%
%%%%%%%%%%%%%%%%%%%%%%%%%%%%%%
\section{Results} \label{sec:results}
Figure~\ref{fig:showline} shows the integrated intensity and intensity-weighted velocity maps of the \co, \tco, and \ceo\ lines. We confirm a bright central north-south bar-like elongated structure, a northern shorter arm and a southern longer arm in all the \co, \tco, and \ceo\ lines, which were previously identified by Submillimeter Array observations \citep{Espada10,Espada18}. Moreover, diffuse flocculent components filling in the inter-arm regions are also identified in \co, which coincide with the diffuse, clumpy H$\alpha$ components \citep{Hattori04}. The \co\ velocity fields show rotating disk-like signature with a ``S"-shape non-circular motion, indicating the presence of inflow motions and/or a warped disk as seen in other galaxies \citep{Espada12,Saito17b}.

Both the \tco\ and \ceo\ images coincide with strong features in the \co\ images. Similar to the asymmetric spiral structures seen in the H$\alpha$, SSC, and CO maps \citep{Espada18}, we found a remarkably thin, long southern arm and an isolated blob at the south part of the northern arm in those isotopologue lines. We note that \ceoone, which is also covered by one of our tunings, was marginally detected only around the very center, and thus we do not utilize this line in this paper.

\subsection{Band~3 and 6 Continuum Emission}
We detected 2.9~mm (Band~3) and 1.3~mm (Band~6) continuum emission toward the central bar-like structure and some parts of the arms in NGC~3110 (Figure~\ref{fig_contin}). For normal star-forming galaxies without strong AGN signature, the 1.3~mm continuum emission is mostly dominated by thermal dust emission whereas thermal free-free emission dominates the 2.9~mm continuum emission with minor contribution from non-thermal synchrotron emission \citep[e.g.,][]{Condon92,Saito16}. We will discuss the physical origin of the emission based on the spectral index in Section~\ref{sec:derive}.

\subsection{Line Intensity Ratios}
\begin{figure*}
\begin{center}
\includegraphics[width=18cm]{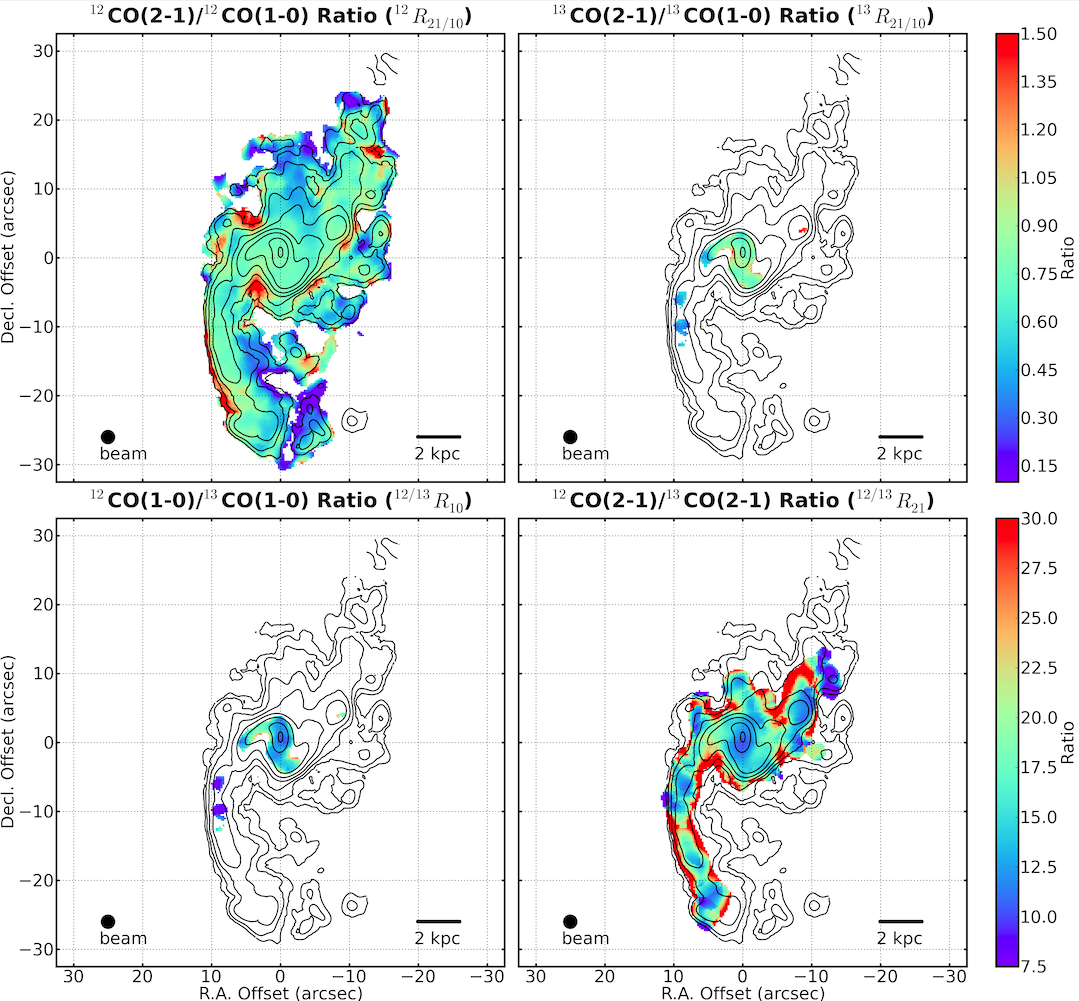}
\end{center}
\caption{(top) \cotwo/\coone\ and \tcotwo/\tcoone\ intensity ratio images of NGC~3110. The contours show the CO~(1--0) integrated intensity image. The center position of this image is the same as Figure~\ref{fig:showline}.
(bottom) Same as the top panel, except for \coone/\tcoone\ and \cotwo/\tcotwo\ intensity ratios.
}\label{fig_ratio}
\end{figure*}

\begin{figure*}
\begin{center}
\includegraphics[width=18cm]{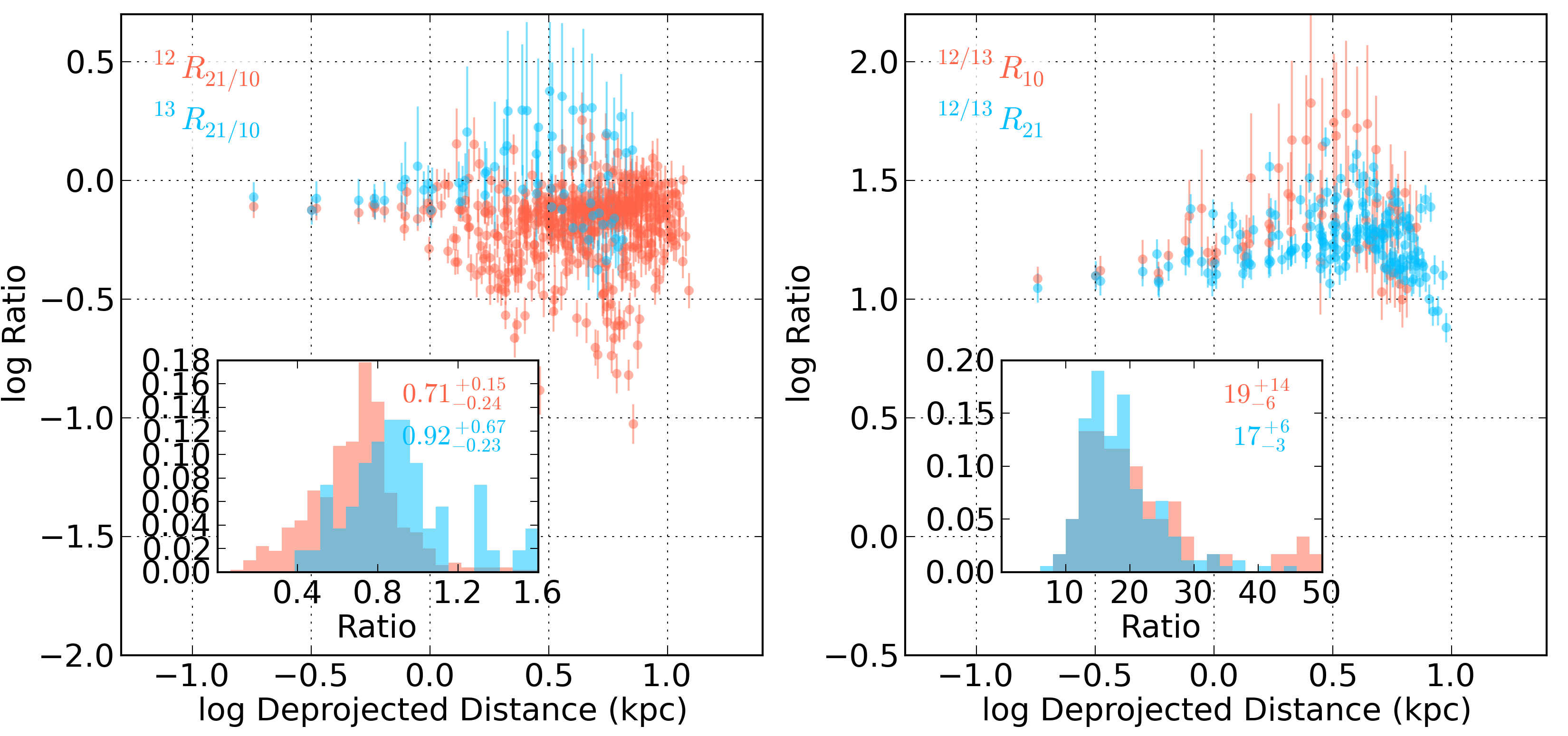}
\end{center}
\caption{(left) The deprojected radial distribution of $^{12}R_{21/10}$ and $^{13}R_{21/10}$. Each data point is measured by the 3\farcs0 ($\sim$1~kpc) aperture. We assumed P.A. = 171\degr\ and inclination = 65\degr\ \citep{Espada18}. The inset shows the ratio distributions. The median values with the 16$^{th}$ and 84$^{th}$ widths are shown at the top-right corner.
(right) The deprojected radial distribution of $^{12/13}R_{10}$ and $^{12/13}R_{21}$.
}\label{fig_radial_ratio}
\end{figure*}

Here we present line ratios made from some combinations of the observed CO lines which are sensitive to galactic scale physical properties of molecular gas ISM. The notations of intensity ratios are defined as,

\begin{eqnarray}
^{X}R_{J_uJ_l/J'_uJ'_l} &\equiv& \frac{I_{X(J_u-J_l)}}{I_{X(J'_u-J'_l)}} \\
^{X/Y}R_{J_uJ_l} &\equiv& \frac{I_{X(J_u-J_l)}}{I_{Y(J_u-J_l)}} \nonumber
\end{eqnarray}
where the first indicates a ratio of different transitions for the main molecule or its isotopologue, and the second indicates a ratio of different isotopes for the same line transition. The spatial distributions of the $J$ $=$ 2--1/$J$ $=$ 1--0 ratios, $^{\rm 12}R_{21/10}$ and $^{\rm 13}R_{21/10}$, are shown in Figure~\ref{fig_ratio} top. The $^{\rm 12}R_{21/10}$ shows a smooth distribution over the entire arms and bar-like elongation ($\sim$0.8 on average). The $^{\rm 13}R_{21/10}$ peaks at the center ($\sim$1.0) with lower values at the ends of the elongation ($\sim$0.7) and the southern arm ($\sim$0.4). The spatial distributions of the \co/\tco\ ratios, $^{\rm 12/13}R_{10}$ and $^{\rm 12/13}R_{21}$, are shown in Figure~\ref{fig_ratio} bottom. Unlike the $^{\rm 12}R_{21/10}$ and $^{\rm 13}R_{21/10}$ distributions, $^{\rm 12/13}R_{21}$ shows lower value ($\sim$10) at the nuclear region, higher values ($\sim$30) toward $\sim$5\arcsec$\:$ east and west from the nucleus, and then lower values again at the tip of the southern arm (Figure~\ref{fig_ratio} bottom). $^{\rm 12/13}R_{10}$ also shows a similar trend although the low signal-to-noise ratio is insufficient.
Those spatial variations can be clearly seen in the radial profiles in Figure~\ref{fig_radial_ratio}.

%%%%%%%%%%%%%%%%%%%%%%%%%%%%%%
%%%%%%%% Derivation %%%%%%%%%%
%%%%%%%%%%%%%%%%%%%%%%%%%%%%%%
\section{Derivation of Physical Parameters} \label{sec:derive}
In this Section, we describe how we derive \aCO, spectral index between 1.3~mm and 2.9~mm, extinction-corrected SFR, and the number density of SSC based on the data taken by using the Nyquist-sampled 3\farcs0 apertures. The derived parameters are visualized in Figure~\ref{fig_Nyquist}.

\subsection{Non-LTE Calculation using RADEX}\label{radex}
We used the non-local thermodynamic equilibrium (LTE) radiative transfer code {\tt RADEX} \citep{vanderTak07} and compare with the observed line ratios. We derived the physical conditions of molecular gas, assuming a single-phase ISM (i.e., the gas physics can be represented by a single set of the excitation parameters), an expanding sphere geometry ($dv$ $=$ 100~\kms), the cosmic microwave background temperature ($T_{\rm bg}$ = $T_{\rm CMB}$ $=$ 2.73K), and [CO]/[H$_2$] ($=$ 3 $\times$ 10$^{-4}$) and [CO]/[\tco] ($=$ 70) abundances which are same as those applied for the LTE calculation described later. The upper state energies and the Einstein coefficients were taken from the {\it Leiden Atomic and Molecular Database} \citep{Schoier05}. We made model grids within a range of the gas kinetic temperature (\Tkin) $=$ 5-205~K, the gas volume density (\nhydro) $=$ 10$^{2.0}$-10$^{7.0}$~\uvolume, and H$_2$ gas column densities (\Nhydro) $=$ 10$^{22.0}$-10$^{22.5}$ cm$^{-2}$. After deriving best-fit values at each aperture via a $\chi^2$ minimization, we made Nyquist-sampled \Tkin\ and \nhydro\ images as shown in Figures~\ref{fig_Nyquist} top-left and top-center. We regard the 90\% confidence area in the parameter space as the uncertainty of the derived \Tkin\ and \nhydro\ following the standard way \citep[e.g.,][]{Sliwa14}.

\subsection{The CO-to-H\texorpdfstring{$_2$}{2} Conversion Factor}
For extragalactic objects, the \coone\ and \cotwo\ lines are popular and well calibrated as H$_2$ gas mass tracers \citep[see][for a review]{Bolatto13}. This is mainly because the sensitivity of the radio instruments is not enough to detect optically-thin low density molecular gas tracers. Recent high-sensitivity interferometric studies provided several ways to calibrate \aCO: gas mass measured by kinematics \citep[i.e., dynamical mass; e.g.,][]{Downes&Solomon98}, gas-to-dust ratio \citep[e.g.,][]{Sandstrom13}, and radiative transfer analysis with multiple molecular lines \citep[e.g.,][]{Sliwa17}.

In this paper, we estimate H$_2$ gas mass ($M_{\rm H_2}$) with two different methods and then compare with $L'_{\rm CO(1-0)}$.

\subsubsection{LTE Mass Derivation}
One reliable way to estimate $M_{\rm H_2}$ is to utilize optically-thin molecular gas tracers. In this paper, we observed two transitions of \tco, which are usually thought to be optically-thin, in order to solve the radiative transfer equation. Multiple transition data allow us to directly estimate the excitation temperature ($T_{\rm ex}$) and the column density of the molecule assuming LTE, resulting in accurate molecular gas mass ($M_{\rm LTE}$). Since the number of apertures with detection of both \tco\ transitions higher than 3$\sigma$ is less than 20 (mostly around the nuclear region and the southern arm), we first determine the typical $T_{\rm ex}$ for those $\lesssim$20 apertures and then assume a single $T_{\rm ex}$ in order to derive the column density for much more apertures. Here we employ the following equation:

\begin{eqnarray}\label{eqn_rot}
\frac{I_{\rm ^{13}CO}}{\rm K\:km\:s^{-1}} &=& \frac{8\pi S\mu_0^2\nu N_{\rm ^{13}CO}}{3kQ(T_{\rm rot})}\left\{1 - \frac{\exp(h\nu/kT_{\rm rot}) - 1}{\exp(h\nu/kT_{\rm bg}) - 1}\right\}\nonumber \\
&&\times\exp\left(-\frac{E_{\rm u}}{kT_{\rm rot}}\right),
\end{eqnarray}
where $S$, $\mu_0$, $\nu$, $N_{\rm ^{13}CO}$, $k$, Q($T_{\rm rot}$), $T_{\rm rot}$, $h$, $T_{\rm bg}$, and $E_{\rm u}$ are line strength, dipole moment, transition frequency, total column density of \tco, the Boltzmann constant, the partition function, rotation temperature which is the same as $T_{\rm ex}$, the Planck constant, the cosmic microwave background temperature, and upper state energy, respectively (e.g., \citealt{Goldsmith99,Watanabe14,Nakajima18}). We took the transition parameters necessary for calculating the equation from Splatalogue\footnote{http://www.cv.nrao.edu/php/splat/} and the {\it Cologne Database for Molecular Spectroscopy} \citep{Muller01,Muller05}.

\begin{figure*}
\begin{center}
\includegraphics[width=18cm]{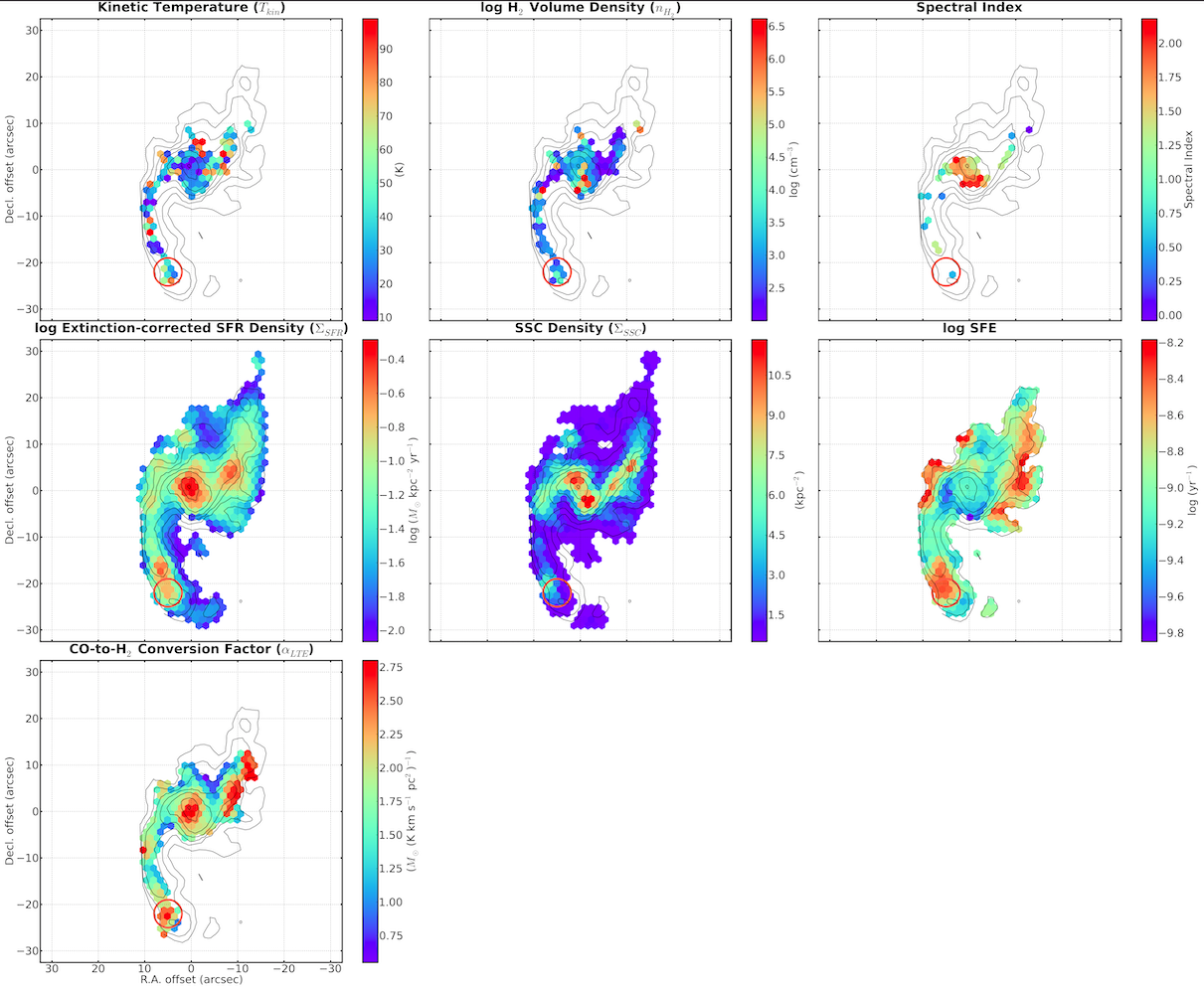}
\end{center}
\caption{(top-left) The \Tkin\ map (RADEX output) of NGC~3110 reconstructed from the 3\farcs0 Nyquist-sampling analysis. The overlaid CO~(1--0) integrated intensity contours are also reconstructed using the same sampling procedure. The red circle presents the approximate position of the region ``A" (see text). The center position of this image is the same as Figure~\ref{fig:showline}.
(top-center) The \nhydro\ map (RADEX output).
(top-right) The 2.9~mm to 1.3~mm spectral index map.
(middle-left) The extinction-corrected SFR surface density (\Ssfr) map in logarithmic scale.
(middle-center) The SSC number density ($\Sigma_{\rm SSC}$) map.
(middle-right) The SFE map in logarithmic scale.
(bottom-left) The CO-to-H$_2$ conversion factor (\aCO) map
 }\label{fig_Nyquist}
\end{figure*}

\begin{figure*}
\begin{center}
\includegraphics[width=18cm]{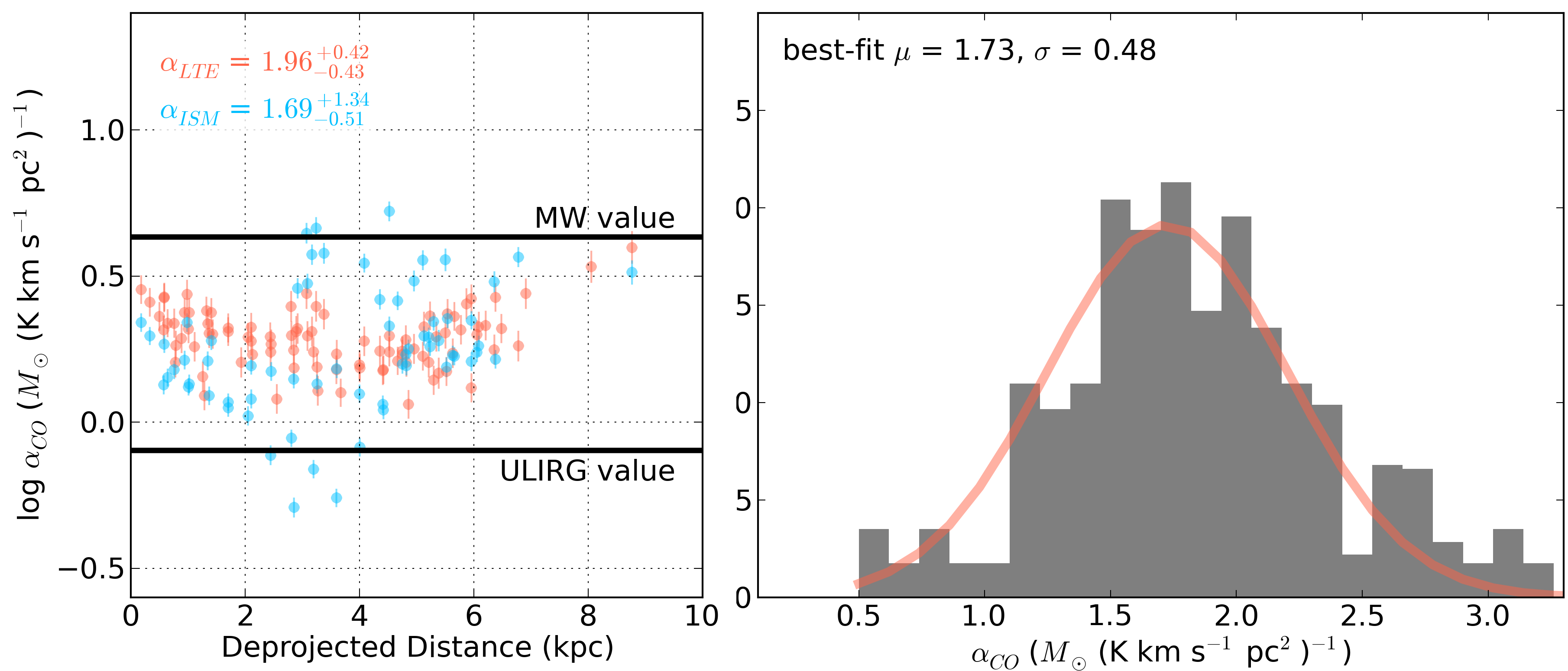}
\end{center}
\caption{
(left) Radial \aCO\ distributions based on the LTE masses and ISM masses. The median values with the 16$^{th}$ and 84$^{th}$ widths are shown at the top-right corner. The Galactic and (U)LIRG values are shown as the black solid lines.
(right) Distribution of the \aCO. The solid red line shows the best-fit Gaussian.
 }\label{fig_alpha}
\end{figure*}

The median and average $T_{\rm rot}$ are 15.1~K and 15.4~K, respectively. Thus we substituted $T_{\rm rot}$ $=$ 15~K for the equation~\ref{eqn_rot} in order to estimate $N_{\rm ^{13}CO}$ for all apertures with \tcotwo\ detection. Then, $M_{\rm LTE}$ is derived by using this equation:
\begin{equation}
\frac{M_{\rm LTE}}{M_{\odot}} = 1.36\frac{N_{\rm ^{13}CO}}{\rm [^{13}CO]/[H_2]}\:m_{\rm H_2}A_{\rm aperture},
\end{equation}
where $m_{\rm H_2}$ is the mass of the hydrogen molecule, $A_{\rm aperture}$ is the area of the apertures, and [\tco]/[H$_2$] is the \tco\ abundance relative to H$_2$. Here we assumed [\tco]/[H$_2$] $=$ 4.3 $\times$ 10$^{-6}$,  that is similar to the standard value observed in Galactic warm, star-forming molecular clouds ([\co]/[\tco] $=$ 70, [\co]/[H$_2$] $=$ 3.0 $\times$ 10$^{-4}$; e.g., \citealt{Blake87,Lacy94}). These are commonly used values in (U)LIRG studies \citep[e.g.,][]{Sliwa14}. In order to accurately constrain the abundance ratios in NGC~3110, we need a more sophisticated modeling with increasing the number of input maps and their quality. We multiply by 1.36 to account for the Helium abundance relative to hydrogen. The derived 16$^{th}$--50$^{th}$--84$^{th}$ percentiles of \aLTE\ ($=$ $M_{\rm LTE}$/$L'_{\rm CO(1-0)}$) are 1.30--1.69--2.21~\uaco. The radial distribution of the derived \aLTE\ with $T_{\rm rot}$ $=$ 15~K is shown as the red data points in Figure~\ref{fig_alpha} left. The spatial distribution is also shown in Figure~\ref{fig_Nyquist} bottom-left.

Here we also derived \aLTE\ using \Tkin\ ($=$ \aLTE(\Tkin)) derived in Section~\ref{radex}, instead of the fixed $T_{\rm rot}$, based on a reasonable assumption that CO excitation is entirely dominated by collision with H$_2$. However, low temperature and low density gas are not usually in thermal equilibrium, so that we expect $T_{\rm rot}$ $<$ \Tkin. Thus, \aLTE(\Tkin) give the upper limit. Indeed, we confirmed that \aLTE(\Tkin) are $\gtrsim$2 times larger than \aLTE($T_{\rm rot}$).

\subsubsection{ISM Mass Derivation}
One of the other methods to estimate the molecular gas mass is to utilize the Rayleigh-Jeans part of the thermal dust continuum emission as described by \citet{Scoville16}. We derived molecular ISM masses ($M_{\rm ISM}$) from the 1.3~mm dust continuum emission using the following equation;
\begin{eqnarray}
\frac{M_{\rm ISM}}{10^{10} M_{\odot}} &=& 1.78\:S_{\rm\nu_{obs}}(1 + z)^{-4.8}\left(\frac{\nu_{\rm obs}}{\nu_{\rm 850\mu m}}\right)^{-3.8}\left(\frac{D_{\rm L}}{\rm Gpc}\right)^2 \nonumber\\
&\times& \left(\frac{6.7 \times 10^{19}}{\alpha_{\rm 850}}\right)\frac{\Gamma_0}{\Gamma_{\rm RJ}},
\end{eqnarray}
where $S_{\rm\nu_{obs}}$ is the observed continuum flux, $\nu_{\rm 850\mu m}$ is 353~GHz, $\nu_{\rm obs}$ is the observed frequency (= 235~GHz), $\alpha_{\rm 850}$ is the calibration constant corresponding to 6.7 $\times$ 10$^{19}$ erg s$^{-1}$ Hz$^{-1}$ $M_{\odot}^{-1}$ introduced by \citet{Scoville16}, and $\Gamma_{\rm RJ}$ and $\Gamma_{\rm 0}$ is given by,
\begin{eqnarray}
\Gamma_{\rm RJ}(T_{\rm d}, \nu_{\rm obs}, z) &=& \frac{h\nu_{\rm obs}(1 + z)/kT_{\rm d}}{e^{h\nu_{\rm obs}(1 + z)/kT_{\rm d}} - 1} \\
\Gamma_{\rm 0} &=& \Gamma_{\rm RJ}(T_{\rm d}, \nu_{\rm 850\mu m}, 0)
\end{eqnarray}
where $T_{\rm d}$ is the dust temperature. Here, we adopt $T_{\rm d}$ $=$ $T_{\rm rot}$ (= 15~K) in order to estimate \aISM\ (= $M_{\rm ISM}$/$L'_{\rm CO(1-0)}$), based on the empirical relationship between molecular gas condition and dust heating \citep{Koda20,Liu21}.

The radial \aISM\ distributions as well as \aLTE\ are shown in Figure~\ref{fig_alpha} left. The derived 16$^{th}$--50$^{th}$--84$^{th}$ percentiles of \aISM\ are 1.08--1.48--2.26~\uaco.

\subsubsection{Appropriate \texorpdfstring{\aCO}{aco} for NGC~3110}\label{alpha}
We found the two methods show a non-monotonic trend (i.e., decreasing then increasing). However, we assume a flat \aCO\ hereafter, as the \aCO\ assumption (e.g., varying or constant) does not affect the discussion of this paper (see Appendix~\ref{App1}). The assumed flat \aCO\ distribution is consistent with spatially-resolved \aCO\ values for other nearby spirals \citep{Sandstrom13} and a LIRG NGC~1614 \citep{Saito17b}. Figure~\ref{fig_alpha} (right) shows the histogram of \aISM($T_{\rm rot}$) with a best-fit Gaussian (peak $=$ 1.73 and dispersion $=$ 0.48). In this paper, we adopt a fixed \aCO\ of 1.7~\uaco\ when measuring gas masses. Note that this \aCO\ value is consistent with values derived for other (U)LIRGs \citep[e.g.,][]{Sliwa17}.

The increasing \aLTE\ trend around the central kpc of NGC~3110 is opposite to the trend found around the center of nearby galaxies \citep{Sandstrom13,Saito17b}. We explain that this increasing trend is due to the assumed constant $T_{\rm rot}$. According to equation~\ref{eqn_rot}, two times higher $T_{\rm rot}$ results in $\lesssim$0.5dex lower $N_{\rm ^{13}CO}$. Thus, the \aLTE\ trend will be flattened or even inversed if we are able to measure $T_{\rm rot}$ at each aperture. Although one of our original motivations to observe multiple transitions is to study the temperature effect on the gas mass measurements around the galaxy centers, the signal-to-noise ratios of our \tcoone\ map is insufficient to measure the spatial $T_{\rm rot}$ distribution.

\subsection{Spectral Index: Two Thermal Components}
The millimeter/submillimeter part of the spectral energy distribution of a star-forming galaxy is known to be dominated by nonthermal (synchrotron) and thermal (free-free and cold dust) emission \citep[e.g.,][]{Condon92}. Especially, the two thermal components are more important at the higher frequency regime ($\gtrsim$100~GHz), because the steep negative spectral index of the nonthermal continuum ($\alpha$ $\sim$ $-$0.8) makes it weaker than the other two ($\alpha$ $\sim$ $-$0.1 and 2-4 for free-free and dust, respectively). Based on our Band~3 and Band~6 data, we can roughly study the continuum property using the following equation:

\begin{equation}
\alpha_{\nu_2}^{\nu_1} = \log_{\frac{\nu_1}{\nu_2}}\frac{S_{\nu_1}}{S_{\nu_2}}
\end{equation}
where $\nu_i$ is the observed frequency, and $S_{\nu_i}$ is the observed flux at $\nu_i$. The derived spectral index image is shown in Figure~\ref{fig_Nyquist} top-right. We found the central part of NGC~3110 is dominated by the steep dust emission, although the southern tip of the southern arm shows relatively low spectral index ($\gtrsim$0.4), implying non-negligible contribution from the free-free emission (i.e., \hii\ regions containing ionizing stars).

\subsection{SFR Derivation and Extinction Correction}
Here we calculate the other important physical parameter, SFR based on previous H$\alpha$ data taken by OAO \citep{Hattori04} and archival 1.4~GHz continuum.

In order to estimate extinction-corrected H$\alpha$ luminosity ($L({\rm H_{\alpha}})_{\rm corr}$), we used the prescription described by \citet{Kennicutt09},
\begin{equation}
\frac{L({\rm H_{\alpha}})_{\rm corr}}{\rm erg\:s^{-1}} = L({\rm H_{\alpha}})_{\rm obs} + 0.39 \times 10^{13}L(\rm 1.4GHz),
\end{equation}
where $L({\rm H_{\alpha}})_{\rm obs}$ is the observed H$\alpha$ luminosity in erg s$^{-1}$ unit and $L(\rm 1.4GHz)$ is the observed 1.4~GHz radio continuum luminosity in erg s$^{-1}$ Hz$^{-1}$ unit. Then, we measure extinction-corrected SFR at each aperture using the following equation (see \citealt{Kennicutt12} for a review),
\begin{equation}
\frac{\rm SFR}{\rm M_{\odot}\:yr^{-1}} = \frac{L({\rm H_{\alpha}})_{\rm corr}}{10^{41.27}}.
\end{equation}
The derived total SFR from the extinction-corrected H$\alpha$ data (i.e., summed over all independent apertures) is $\sim$18~\usfr (Figure~\ref{fig_Nyquist} middle-left).

We caution that the employed extinction-correction prescription is based on the global measurements (see \citealt{Kennicutt09}), and therefore our spatially-resolved SFRs are uncertain. However, we confirmed that the measured total SFR is in agreement with values derived in the literature: \citet{U12} derived $\sim$24~\usfr\ based on the IR SED, and \citet{Espada18} derived 19.7~\usfr\ and 21.2~\usfr\ based on H$\alpha$+24~$\mu$m and 24~$\mu$m, respectively. In addition, the global trend seen in the Kennicutt-Schmidt (KS) relation (see next Section) is also consistent with the relation described in \citet{Espada18} (although the scatter is larger in our KS relation, which is clearly seen in data points with the deprojected distance from 3 to 6.5~kpc). After these double-checking, we decided to use this prescription in this paper. Constructing an accurate SFR map at high angular resolution comparable to the ALMA resolution is beyond the scope of this paper, but a necessary step to further understanding the star formation activities in NGC~3110.

\begin{figure*}
\begin{center}
\includegraphics[width=15cm]{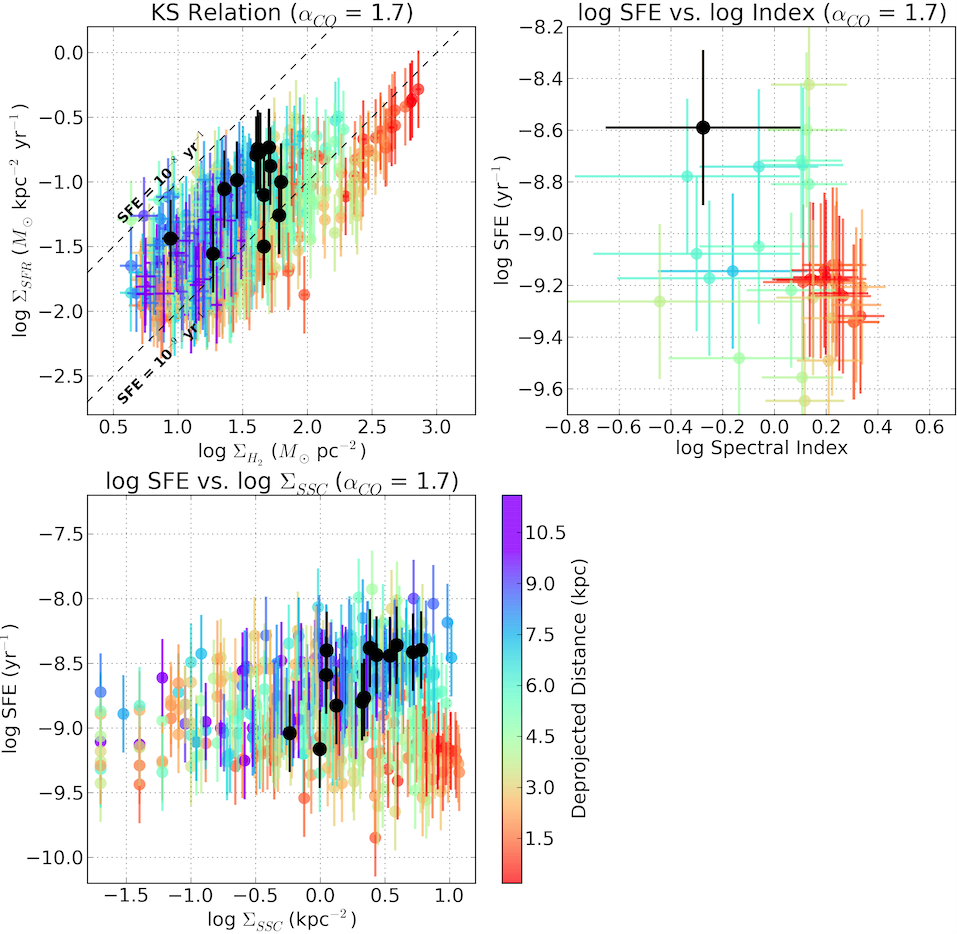}
\end{center}
\caption{(top-left) The 1 kpc-scale Kennicutt-Schmidt relation of NGC~3110. The black data indicate measurements toward the region ``A". The dashed lines show constant SFE of 10$^{-8}$ and 10$^{-9}$~yr$^{-1}$. The color scale corresponds to the distance from the nucleus.
(top-right) SFE plots against spectral index.
(bottom) SFE plots against $\Sigma_{\rm SSC}$.
}\label{fig_sfe}
\end{figure*}

\subsection{Kennicutt-Schmidt Relation}
We plot the ($\sim$1~kpc-scale) NGC~3110 data points on the Kennicutt-Schmidt \Ssfr\ - \Shydro\ relation \citep{Kennicutt98b} as shown in Figure~\ref{fig_sfe} top-left. Note that if adopting variable \aLTE\ the KS relation does not qualitatively change, and even the difference between the circumnuclear region and the arms becomes larger; i.e., the central part of this galaxy shows higher densities, and SFE ($=$ \Ssfr/\Shydro) is high 7-8 kpc away from the nucleus. The SFE map using a fixed \aCO\ is shown in Figure~\ref{fig_Nyquist} middle-right. The highest SFE can be found inside the red circle around the southern tip of the southern arm.

Comparing with similar resolution ($\sim$0.75~kpc) data for local non-interacting spiral galaxies \citep{Bigiel08}, NGC~3110 shows comparable or higher \Ssfr\ and \Shydro\ than spiral galaxies with highest \Ssfr\ ($\lesssim$1~\udsfr) and \Shydro\ ($\lesssim$10$^{2.5}$~\udhydro) even though NGC~3110 is an early stage of interaction. The SFEs of NGC~3110 range from 10$^{-8}$ to 10$^{-9}$~yr$^{-1}$, which is higher than that of spiral galaxies ($\sim$5 $\times$ 10$^{-10}$~yr$^{-1}$). Especially, at the southern tip of the southern arm (hereafter region ``A"; a red circle in Figure~\ref{fig_Nyquist}), the SFE is $\sim$10$^{-8}$~yr$^{-1}$, which is $\sim$4 times higher than the average SFE around the nuclear region. Our SFE values are consistent with previous SFE values \citep{Espada18} if we use the same \aCO\ they assumed (i.e., 4.3~\uaco instead of 1.7~\uaco).

We note that the prescriptions to derive SFR are different among this study and other papers introduced in this Section; far-ultraviolet and 24$\mu$m in \citet{Bigiel08}, H$\alpha$ and 24$\mu$m in \citet{Espada18}, and H$\alpha$ and radio continuum in this study. This introduces systematic uncertainties in the absolute values, and thus hereafter we focus on our own data.

%%%%%%%%%%%%%%%%%%%%%%%%%%%%%%
%%%%%%%%% Discussion %%%%%%%%%
%%%%%%%%%%%%%%%%%%%%%%%%%%%%%%
\section{Discussion} \label{sec:discussion}
\subsection{Star Formation at the Tip of the Southern Arm}
The southern tip of the southern arm, region ``A", shows the highest SFE among the NGC~3110 data points, which is an unusual feature for spiral galaxies which typically show highest SFE around the center \citep{Leroy08}. Here we try to characterize the physical properties of the highest SFE region.

One of the intriguing features seen in the region ``A" is the shallow spectral index between 1.3~mm and 2.9~mm. In Figure~\ref{fig_sfe} top-right, we plot the SFE against the spectral index. The data points with the deprojected radius $<$ 3~kpc (yellow to red color) tend to distribute around the spectral index $\sim$1--3, indicating dusty star formation at the central region of NGC~3110. In contrast to the center, the outer part, especially the region ``A", shows much lower spectral index ($\sim$0.5), implying that 2.9~mm continuum from the region ``A" is likely to be significantly contaminated by the free-free emission from ionized gas by young massive stars (typical age $\lesssim$ 10~Myr; \citealt{Kennicutt12}). Such ionized gas can be also traced by H$\alpha$ emission, which is used to derive the SFE. Considering that infrared continuum emission (3-1100~$\mu$m) is a good tracer of star formation with the age $\lesssim$~100~Myr \citep{Kennicutt12}, the trend seen in Figure~\ref{fig_sfe} top-right tells us the dominant age of the star-forming regions. Thus, the region ``A" is likely to be an active and young star-forming region in the disk of NGC~3110, whereas the central part of NGC~3110 is a site of more dusty, long-continued star formation. We note that the discussion of the spectral index in the region ``A" is based on a single data point as shown in the top-right panel of Figure~\ref{fig_sfe}. Although the S/N ratio is sufficient to safely conclude the relatively lower spectral index in the region ``A", higher accuracy continuum observations are required to better constrain age differences.

We plot SFE against $\Sigma_{\rm SSC}$ in Figure~\ref{fig_sfe} bottom, showing no clear correlation between them. However, it has a tendency that the inner regions (i.e., redder points) show relatively lower SFE ($\gtrsim$10$^9$~yr$^{-1}$) but higher $\Sigma_{\rm SSC}$ ($\sim$10~kpc$^{-2}$). Since the typical age of SSCs is thought to be 10-100~Myr \citep{Randriamanakoto15}, the trend seen in Figure~\ref{fig_sfe} bottom supports the idea that region ``A" is younger massive star-forming region compared with those around the center.
% This trend can be also interpreted by the age of the star-forming regions.

\subsection{Early Stage Minor Merger as a Trigger of Wide-spread Star Formation}
The relative age difference between the star-forming regions around the central region and region ``A" is an important clue to understand the physical origin of the kpc-scale star formation in the disk of NGC~3110. Numerical simulations suggested that tidal interaction between large mass ratio galaxies enhances both a long-lived $m =$ 2 bar and transient asymmetric arms \citep{Iono04,Cox08,Pettitt18}. \citet{Espada18} reproduced the morphological characteristics of NGC~3110 using hydrodynamical simulations, and found that NGC~3110 might be experiencing one of the highest SFR epoch during the merger history before coalescence. The typical SSC age is comparable or shorter than the timescale of the merger-induced asymmetric arms in NGC~3110 reproduced by this simulation, which is consistent with the picture that the star-forming region in the region ``A" is triggered by tidal interaction. The star-forming regions triggered on the tidally-induced transient arms show clumpy structures in the simulations as we saw in the CO isotope images of NGC~3110 (Figure~\ref{fig:showline}). Based on those morphological similarities between NGC~3110 and numerical minor merger simulations, we suggest that the on-going interaction with the smaller companion galaxy MCG-01-26-013 triggers an active, young star-forming region (SFE $\sim$ 10$^{-8.2}$~yr$^{-1}$, $\Ssfr$ $\sim$ 10$^{-0.6}$\udsfr, $\Sssc$ $\sim$ 6.0~kpc$^{-2}$) at the southern tip of the southern arm in NGC~3110. The elevated star-forming activities seen in the center of NGC~3110 might be due to merger-driven tidal torques leading to rapid, large-scale gas inflow.

%%%%%%%%%%%%%%%%%%%%%%%%%%%%%%
%%%%%%%%%% Summary %%%%%%%%%%%
%%%%%%%%%%%%%%%%%%%%%%%%%%%%%%
\section{Summary} \label{sec:summary}
We present $\sim$1\farcs0 resolution ALMA Band~3 and Band~6 observations of the rotational transitions of CO and its isotopologues and continuum emission in the nearby interacting LIRG NGC~3110. NGC~3110 has an asymmetric two-armed spirals and a bar-like structure, both of which are likely to be enhanced during the early stage galaxy merger interaction as suggested by previous molecular gas observations and numerical simulations \citep{Espada18,Pettitt18}. Combining our new ALMA datasets with previous H$\alpha$ and 1.4~GHz continuum data, we find that the highest SFE region is located at the southern tip of the southern spiral arm (region ``A"). The region ``A" is characterized with (1) lower spectral index between 1.3~mm and 2.9~mm ($\lesssim$0.5), i.e., a dominant contribution from free-free emission, (2) high number density of super star clusters ($\Sssc$ $\sim$ 6.0~kpc$^{-2}$), and (3) relatively active star formation (SFE $\sim$ 10$^{-8.2}$~yr$^{-1}$ and $\Ssfr$ $\sim$ 10$^{-0.6}$\udsfr). Those observational evidences support the idea that the region ``A" is active, young star-forming region found in NGC~3110. With the strong similarities of the observed gas and star formation properties with numerical minor merger simulations, we suggest that the on-going interaction with a companion galaxy MCG-01-26-013 is the main driver of the star formation at the region ``A" as well as the central starburst activities.

To further investigate the physical properties and the origin of the star-forming regions in the disk of NGC~3110, it is essential to understand giant molecular cloud properties, i.e., a site of star formation, which can be addressed by current long baseline capability of ALMA.

%%%%%%%%%%%%%%%%%%%%%%%%%%%%%%
%%%%%% Acknowledgments %%%%%%%
%%%%%%%%%%%%%%%%%%%%%%%%%%%%%%
\begin{acknowledgments}
Y.K., T.S., and other authors thank ALMA staff for their kind support. This work was supported by the ALMA Japan Research Grant of NAOJ Chile Observatory, NAOJ-ALMA-0114. DE acknowledges support from a Beatriz Galindo senior fellowship (BG20/00224) from the Spanish Ministry of Science and Innovation. This paper makes use of the following ALMA data: ADS/JAO.ALMA\#2013.0.01172.S. ALMA is a partnership of ESO (representing its member states), NSF (USA) and NINS (Japan), together with NRC (Canada), MOST and ASIAA (Taiwan), and KASI (Republic of Korea), in cooperation with the Republic of Chile. The Joint ALMA Observatory is operated by ESO, AUI/NRAO and NAOJ. This NVAS image was produced as part of the NRAO VLA Archive Survey, (c) AUI/NRAO. This research has made use of the NASA/IPAC Extragalactic Database (NED), which is funded by the National Aeronautics and Space Administration and operated by the California Institute of Technology. This work is based in part on observations made with the {\it Spitzer} Space Telescope, which is operated by the Jet Propulsion Laboratory, California Institute of Technology under a contract with NASA.
\end{acknowledgments}

\vspace{5mm}
%\facilities{ALMA, VLA, VLT(MUSE)}

\software{
{\tt ALMA Calibration Pipeline},
{\tt astropy} \citep{Astropy13,Astropy18},
{\tt CASA} \citep{McMullin07},
{\tt matplotlib} \citep{Hunter07},
{\tt numpy} \citep{Oliphant06},
{\tt scipy} \citep{Virtanen20}
}

%%%%%%%%%%%%%%%%%%%%%%%%%%%%%%
%%%%%%%% bibliography %%%%%%%%
%%%%%%%%%%%%%%%%%%%%%%%%%%%%%%
\bibliography{n3110_co}{}

\begin{thebibliography}{}
\expandafter\ifx\csname natexlab\endcsname\relax\def\natexlab#1{#1}\fi
\providecommand{\url}[1]{\href{#1}{#1}}
\providecommand{\dodoi}[1]{doi:~\href{http://doi.org/#1}{\nolinkurl{#1}}}
\providecommand{\doeprint}[1]{\href{http://ascl.net/#1}{\nolinkurl{http://ascl.net/#1}}}
\providecommand{\doarXiv}[1]{\href{https://arxiv.org/abs/#1}{\nolinkurl{https://arxiv.org/abs/#1}}}

\bibitem[{{Armus} {et~al.}(2009){Armus}, {Mazzarella}, {Evans}, {Surace},
  {Sanders}, {Iwasawa}, {Frayer}, {Howell}, {Chan}, {Petric}, {Vavilkin},
  {Kim}, {Haan}, {Inami}, {Murphy}, {Appleton}, {Barnes}, {Bothun}, {Bridge},
  {Charmandaris}, {Jensen}, {Kewley}, {Lord}, {Madore}, {Marshall},
  {Melbourne}, {Rich}, {Satyapal}, {Schulz}, {Spoon}, {Sturm}, {U}, {Veilleux},
  \& {Xu}}]{Armus09}
{Armus}, L., {Mazzarella}, J.~M., {Evans}, A.~S., {et~al.} 2009, \pasp, 121,
  559, \dodoi{10.1086/600092}

\bibitem[{{Astropy Collaboration} {et~al.}(2013){Astropy Collaboration},
  {Robitaille}, {Tollerud}, {Greenfield}, {Droettboom}, {Bray}, {Aldcroft},
  {Davis}, {Ginsburg}, {Price-Whelan}, {Kerzendorf}, {Conley}, {Crighton},
  {Barbary}, {Muna}, {Ferguson}, {Grollier}, {Parikh}, {Nair}, {Unther},
  {Deil}, {Woillez}, {Conseil}, {Kramer}, {Turner}, {Singer}, {Fox}, {Weaver},
  {Zabalza}, {Edwards}, {Azalee Bostroem}, {Burke}, {Casey}, {Crawford},
  {Dencheva}, {Ely}, {Jenness}, {Labrie}, {Lim}, {Pierfederici}, {Pontzen},
  {Ptak}, {Refsdal}, {Servillat}, \& {Streicher}}]{Astropy13}
{Astropy Collaboration}, {Robitaille}, T.~P., {Tollerud}, E.~J., {et~al.} 2013,
  \aap, 558, A33, \dodoi{10.1051/0004-6361/201322068}

\bibitem[{{Astropy Collaboration} {et~al.}(2018){Astropy Collaboration},
  {Price-Whelan}, {Sip{\H{o}}cz}, {G{\"u}nther}, {Lim}, {Crawford}, {Conseil},
  {Shupe}, {Craig}, {Dencheva}, {Ginsburg}, {VanderPlas}, {Bradley},
  {P{\'e}rez-Su{\'a}rez}, {de Val-Borro}, {Aldcroft}, {Cruz}, {Robitaille},
  {Tollerud}, {Ardelean}, {Babej}, {Bach}, {Bachetti}, {Bakanov}, {Bamford},
  {Barentsen}, {Barmby}, {Baumbach}, {Berry}, {Biscani}, {Boquien}, {Bostroem},
  {Bouma}, {Brammer}, {Bray}, {Breytenbach}, {Buddelmeijer}, {Burke},
  {Calderone}, {Cano Rodr{\'\i}guez}, {Cara}, {Cardoso}, {Cheedella}, {Copin},
  {Corrales}, {Crichton}, {D'Avella}, {Deil}, {Depagne}, {Dietrich}, {Donath},
  {Droettboom}, {Earl}, {Erben}, {Fabbro}, {Ferreira}, {Finethy}, {Fox},
  {Garrison}, {Gibbons}, {Goldstein}, {Gommers}, {Greco}, {Greenfield},
  {Groener}, {Grollier}, {Hagen}, {Hirst}, {Homeier}, {Horton}, {Hosseinzadeh},
  {Hu}, {Hunkeler}, {Ivezi{\'c}}, {Jain}, {Jenness}, {Kanarek}, {Kendrew},
  {Kern}, {Kerzendorf}, {Khvalko}, {King}, {Kirkby}, {Kulkarni}, {Kumar},
  {Lee}, {Lenz}, {Littlefair}, {Ma}, {Macleod}, {Mastropietro}, {McCully},
  {Montagnac}, {Morris}, {Mueller}, {Mumford}, {Muna}, {Murphy}, {Nelson},
  {Nguyen}, {Ninan}, {N{\"o}the}, {Ogaz}, {Oh}, {Parejko}, {Parley}, {Pascual},
  {Patil}, {Patil}, {Plunkett}, {Prochaska}, {Rastogi}, {Reddy Janga},
  {Sabater}, {Sakurikar}, {Seifert}, {Sherbert}, {Sherwood-Taylor}, {Shih},
  {Sick}, {Silbiger}, {Singanamalla}, {Singer}, {Sladen}, {Sooley},
  {Sornarajah}, {Streicher}, {Teuben}, {Thomas}, {Tremblay}, {Turner},
  {Terr{\'o}n}, {van Kerkwijk}, {de la Vega}, {Watkins}, {Weaver}, {Whitmore},
  {Woillez}, {Zabalza}, \& {Astropy Contributors}}]{Astropy18}
{Astropy Collaboration}, {Price-Whelan}, A.~M., {Sip{\H{o}}cz}, B.~M., {et~al.}
  2018, \aj, 156, 123, \dodoi{10.3847/1538-3881/aabc4f}

\bibitem[{{Barnes} \& {Hernquist}(1992)}]{Barnes92}
{Barnes}, J.~E., \& {Hernquist}, L. 1992, \araa, 30, 705,
  \dodoi{10.1146/annurev.aa.30.090192.003421}

\bibitem[{{Bigiel} {et~al.}(2008){Bigiel}, {Leroy}, {Walter}, {Brinks}, {de
  Blok}, {Madore}, \& {Thornley}}]{Bigiel08}
{Bigiel}, F., {Leroy}, A., {Walter}, F., {et~al.} 2008, \aj, 136, 2846,
  \dodoi{10.1088/0004-6256/136/6/2846}

\bibitem[{{Blake} {et~al.}(1987){Blake}, {Sutton}, {Masson}, \&
  {Phillips}}]{Blake87}
{Blake}, G.~A., {Sutton}, E.~C., {Masson}, C.~R., \& {Phillips}, T.~G. 1987,
  \apj, 315, 621, \dodoi{10.1086/165165}

\bibitem[{{Bolatto} {et~al.}(2013){Bolatto}, {Wolfire}, \& {Leroy}}]{Bolatto13}
{Bolatto}, A.~D., {Wolfire}, M., \& {Leroy}, A.~K. 2013, \araa, 51, 207,
  \dodoi{10.1146/annurev-astro-082812-140944}

\bibitem[{{Condon}(1992)}]{Condon92}
{Condon}, J.~J. 1992, \araa, 30, 575,
  \dodoi{10.1146/annurev.aa.30.090192.003043}

\bibitem[{{Cornwell}(2008)}]{Cornwell08}
{Cornwell}, T.~J. 2008, IEEE Journal of Selected Topics in Signal Processing,
  2, 793, \dodoi{10.1109/JSTSP.2008.2006388}

\bibitem[{{Cox} {et~al.}(2008){Cox}, {Jonsson}, {Somerville}, {Primack}, \&
  {Dekel}}]{Cox08}
{Cox}, T.~J., {Jonsson}, P., {Somerville}, R.~S., {Primack}, J.~R., \& {Dekel},
  A. 2008, \mnras, 384, 386, \dodoi{10.1111/j.1365-2966.2007.12730.x}

\bibitem[{{den Brok} {et~al.}(2021){den Brok}, {Chatzigiannakis}, {Bigiel},
  {Puschnig}, {Barnes}, {Leroy}, {Jim{\'e}nez-Donaire}, {Usero}, {Schinnerer},
  {Rosolowsky}, {Faesi}, {Grasha}, {Hughes}, {Kruijssen}, {Liu}, {Neumann},
  {Pety}, {Querejeta}, {Saito}, {Schruba}, \& {Stuber}}]{denBrok21}
{den Brok}, J.~S., {Chatzigiannakis}, D., {Bigiel}, F., {et~al.} 2021, \mnras,
  504, 3221, \dodoi{10.1093/mnras/stab859}

\bibitem[{{Downes} \& {Solomon}(1998)}]{Downes&Solomon98}
{Downes}, D., \& {Solomon}, P.~M. 1998, \apj, 507, 615, \dodoi{10.1086/306339}

\bibitem[{{Elmegreen} {et~al.}(2017){Elmegreen}, {Elmegreen}, {Kaufman},
  {Brinks}, {Struck}, {Bournaud}, {Sheth}, \& {Juneau}}]{Elmegreen17}
{Elmegreen}, D.~M., {Elmegreen}, B.~G., {Kaufman}, M., {et~al.} 2017, \apj,
  841, 43, \dodoi{10.3847/1538-4357/aa6ba5}

\bibitem[{{Espada} {et~al.}(2012){Espada}, {Matsushita}, {Peck}, {Henkel},
  {Israel}, \& {Iono}}]{Espada12}
{Espada}, D., {Matsushita}, S., {Peck}, A.~B., {et~al.} 2012, \apjl, 756, L10,
  \dodoi{10.1088/2041-8205/756/1/L10}

\bibitem[{{Espada} {et~al.}(2010){Espada}, {Martin}, {Hsieh}, {Ho},
  {Matsushita}, {Verdes-Montenegro}, {Sabater}, {Verley}, {Krips}, \&
  {Espigares}}]{Espada10}
{Espada}, D., {Martin}, S., {Hsieh}, P.~Y., {et~al.} 2010, in Galaxies and
  their Masks, ed. D.~L. {Block}, K.~C. {Freeman}, \& I.~{Puerari}, 97,
  \dodoi{10.1007/978-1-4419-7317-7\_7}

\bibitem[{{Espada} {et~al.}(2018){Espada}, {Martin}, {Verley}, {Pettitt},
  {Matsushita}, {Argudo-Fern{\'a}ndez}, {Randriamanakoto}, {Hsieh}, {Saito},
  {Miura}, {Kawana}, {Sabater}, {Verdes-Montenegro}, {Ho}, {Kawabe}, \&
  {Iono}}]{Espada18}
{Espada}, D., {Martin}, S., {Verley}, S., {et~al.} 2018, \apj, 866, 77,
  \dodoi{10.3847/1538-4357/aae07e}

\bibitem[{{Goldsmith} \& {Langer}(1999)}]{Goldsmith99}
{Goldsmith}, P.~F., \& {Langer}, W.~D. 1999, \apj, 517, 209,
  \dodoi{10.1086/307195}

\bibitem[{{Hainline} {et~al.}(2004){Hainline}, {Scoville}, {Yun}, {Hawkins},
  {Frayer}, \& {Isaak}}]{Hainline04}
{Hainline}, L.~J., {Scoville}, N.~Z., {Yun}, M.~S., {et~al.} 2004, \apj, 609,
  61, \dodoi{10.1086/420920}

\bibitem[{{Hattori} {et~al.}(2004){Hattori}, {Yoshida}, {Ohtani}, {Sugai},
  {Ishigaki}, {Sasaki}, {Hayashi}, {Ozaki}, {Ishii}, \& {Kawai}}]{Hattori04}
{Hattori}, T., {Yoshida}, M., {Ohtani}, H., {et~al.} 2004, \aj, 127, 736,
  \dodoi{10.1086/381060}

\bibitem[{{Hubble}(1926)}]{Hubble26}
{Hubble}, E.~P. 1926, \apj, 64, 321, \dodoi{10.1086/143018}

\bibitem[{Hunter(2007)}]{Hunter07}
Hunter, J.~D. 2007, Computing in Science \& Engineering, 9, 90,
  \dodoi{10.1109/MCSE.2007.55}

\bibitem[{{Iono} {et~al.}(2004){Iono}, {Yun}, \& {Mihos}}]{Iono04}
{Iono}, D., {Yun}, M.~S., \& {Mihos}, J.~C. 2004, \apj, 616, 199,
  \dodoi{10.1086/424797}

\bibitem[{{Iono} {et~al.}(2013){Iono}, {Saito}, {Yun}, {Kawabe}, {Espada},
  {Hagiwara}, {Imanishi}, {Izumi}, {Kohno}, {Motohara}, {Nakanishi}, {Sugai},
  {Tateuchi}, {Tamura}, {Ueda}, \& {Yoshii}}]{Iono13}
{Iono}, D., {Saito}, T., {Yun}, M.~S., {et~al.} 2013, \pasj, 65, L7,
  \dodoi{10.1093/pasj/65.3.L7}

\bibitem[{{Kaneko} {et~al.}(2018){Kaneko}, {Kuno}, \& {Saitoh}}]{Kaneko18}
{Kaneko}, H., {Kuno}, N., \& {Saitoh}, T.~R. 2018, \apjl, 860, L14,
  \dodoi{10.3847/2041-8213/aac895}

\bibitem[{{Kennicutt}(1998{\natexlab{a}})}]{Kennicutt98a}
{Kennicutt}, Robert~C., J. 1998{\natexlab{a}}, \araa, 36, 189,
  \dodoi{10.1146/annurev.astro.36.1.189}

\bibitem[{{Kennicutt}(1998{\natexlab{b}})}]{Kennicutt98b}
---. 1998{\natexlab{b}}, \apj, 498, 541, \dodoi{10.1086/305588}

\bibitem[{{Kennicutt} {et~al.}(2009){Kennicutt}, {Hao}, {Calzetti},
  {Moustakas}, {Dale}, {Bendo}, {Engelbracht}, {Johnson}, \&
  {Lee}}]{Kennicutt09}
{Kennicutt}, Robert~C., J., {Hao}, C.-N., {Calzetti}, D., {et~al.} 2009, \apj,
  703, 1672, \dodoi{10.1088/0004-637X/703/2/1672}

\bibitem[{{Kennicutt} \& {Evans}(2012)}]{Kennicutt12}
{Kennicutt}, R.~C., \& {Evans}, N.~J. 2012, \araa, 50, 531,
  \dodoi{10.1146/annurev-astro-081811-125610}

\bibitem[{{Koda} {et~al.}(2020){Koda}, {Sawada}, {Sakamoto}, {Hirota}, {Egusa},
  {Boissier}, {Calzetti}, {Meyer}, {Elmegreen}, {de Paz}, {Harada}, {Ho},
  {Kobayashi}, {Kuno}, {Mart{\'\i}n}, {Muraoka}, {Nakanishi}, {Scoville},
  {Seibert}, {Vlahakis}, \& {Watanabe}}]{Koda20}
{Koda}, J., {Sawada}, T., {Sakamoto}, K., {et~al.} 2020, \apjl, 890, L10,
  \dodoi{10.3847/2041-8213/ab70b7}

\bibitem[{{Lacy} {et~al.}(1994){Lacy}, {Knacke}, {Geballe}, \&
  {Tokunaga}}]{Lacy94}
{Lacy}, J.~H., {Knacke}, R., {Geballe}, T.~R., \& {Tokunaga}, A.~T. 1994,
  \apjl, 428, L69, \dodoi{10.1086/187395}

\bibitem[{{Leroy} {et~al.}(2008){Leroy}, {Walter}, {Brinks}, {Bigiel}, {de
  Blok}, {Madore}, \& {Thornley}}]{Leroy08}
{Leroy}, A.~K., {Walter}, F., {Brinks}, E., {et~al.} 2008, \aj, 136, 2782,
  \dodoi{10.1088/0004-6256/136/6/2782}

\bibitem[{{Liu} {et~al.}(2021){Liu}, {Daddi}, {Schinnerer}, {Saito}, {Leroy},
  {Silverman}, {Valentino}, {Magdis}, {Gao}, {Jin}, {Puglisi}, \&
  {Groves}}]{Liu21}
{Liu}, D., {Daddi}, E., {Schinnerer}, E., {et~al.} 2021, \apj, 909, 56,
  \dodoi{10.3847/1538-4357/abd801}

\bibitem[{Lundgren(2013)}]{Lundgren13}
Lundgren, A. 2013, {ALMA Cycle 2 Technical Handbook Version 1.1}, ALMA.
\newblock
  \url{https://arc.iram.fr/documents/cycle2/alma-technical-handbook.pdf}

\bibitem[{{McMullin} {et~al.}(2007){McMullin}, {Waters}, {Schiebel}, {Young},
  \& {Golap}}]{McMullin07}
{McMullin}, J.~P., {Waters}, B., {Schiebel}, D., {Young}, W., \& {Golap}, K.
  2007, in Astronomical Society of the Pacific Conference Series, Vol. 376,
  Astronomical Data Analysis Software and Systems XVI, ed. R.~A. {Shaw},
  F.~{Hill}, \& D.~J. {Bell}, 127

\bibitem[{{Mihos} \& {Hernquist}(1996)}]{Mihos96}
{Mihos}, J.~C., \& {Hernquist}, L. 1996, \apj, 464, 641, \dodoi{10.1086/177353}

\bibitem[{{M{\"u}ller} {et~al.}(2005){M{\"u}ller}, {Schl{\"o}der}, {Stutzki},
  \& {Winnewisser}}]{Muller05}
{M{\"u}ller}, H. S.~P., {Schl{\"o}der}, F., {Stutzki}, J., \& {Winnewisser}, G.
  2005, Journal of Molecular Structure, 742, 215,
  \dodoi{10.1016/j.molstruc.2005.01.027}

\bibitem[{{M{\"u}ller} {et~al.}(2001){M{\"u}ller}, {Thorwirth}, {Roth}, \&
  {Winnewisser}}]{Muller01}
{M{\"u}ller}, H.~S.~P., {Thorwirth}, S., {Roth}, D.~A., \& {Winnewisser}, G.
  2001, \aap, 370, L49, \dodoi{10.1051/0004-6361:20010367}

\bibitem[{{Nakajima} {et~al.}(2018){Nakajima}, {Takano}, {Kohno}, {Harada}, \&
  {Herbst}}]{Nakajima18}
{Nakajima}, T., {Takano}, S., {Kohno}, K., {Harada}, N., \& {Herbst}, E. 2018,
  \pasj, 70, 7, \dodoi{10.1093/pasj/psx153}

\bibitem[{Oliphant(2006)}]{Oliphant06}
Oliphant, T. 2006, {NumPy}: A guide to {NumPy}, USA: Trelgol Publishing.
\newblock \url{http://www.numpy.org/}

\bibitem[{{Pettitt} {et~al.}(2017){Pettitt}, {Tasker}, {Wadsley}, {Keller}, \&
  {Benincasa}}]{Pettitt17}
{Pettitt}, A.~R., {Tasker}, E.~J., {Wadsley}, J.~W., {Keller}, B.~W., \&
  {Benincasa}, S.~M. 2017, \mnras, 468, 4189, \dodoi{10.1093/mnras/stx736}

\bibitem[{{Pettitt} \& {Wadsley}(2018)}]{Pettitt18}
{Pettitt}, A.~R., \& {Wadsley}, J.~W. 2018, \mnras, 474, 5645,
  \dodoi{10.1093/mnras/stx3129}

\bibitem[{{Randriamanakoto}(2015)}]{Randriamanakoto15}
{Randriamanakoto}, Z. 2015, PhD thesis, Department of Astronomy, University of
  Cape Town, Private Bag X3, Rondebosch 7701, South Africa

\bibitem[{{Randriamanakoto} \& {V{\"a}is{\"a}nen}(2017)}]{Randriamanakoto17}
{Randriamanakoto}, Z., \& {V{\"a}is{\"a}nen}, P. 2017, in Formation, Evolution,
  and Survival of Massive Star Clusters, ed. C.~{Charbonnel} \& A.~{Nota}, Vol.
  316, 70--76, \dodoi{10.1017/S1743921315010510}

\bibitem[{{Randriamanakoto} {et~al.}(2013){Randriamanakoto},
  {V{\"a}is{\"a}nen}, {Ryder}, {Kankare}, {Kotilainen}, \&
  {Mattila}}]{Randriamanakoto13}
{Randriamanakoto}, Z., {V{\"a}is{\"a}nen}, P., {Ryder}, S., {et~al.} 2013,
  \mnras, 431, 554, \dodoi{10.1093/mnras/stt185}

\bibitem[{{Randriamanakoto} {et~al.}(2019){Randriamanakoto},
  {V{\"a}is{\"a}nen}, {Ryder}, \& {Ranaivomanana}}]{Randriamanakoto19}
{Randriamanakoto}, Z., {V{\"a}is{\"a}nen}, P., {Ryder}, S.~D., \&
  {Ranaivomanana}, P. 2019, \mnras, 482, 2530, \dodoi{10.1093/mnras/sty2837}

\bibitem[{{Saito} {et~al.}(2015){Saito}, {Iono}, {Yun}, {Ueda}, {Nakanishi},
  {Sugai}, {Espada}, {Imanishi}, {Motohara}, {Hagiwara}, {Tateuchi}, {Lee}, \&
  {Kawabe}}]{Saito15}
{Saito}, T., {Iono}, D., {Yun}, M.~S., {et~al.} 2015, \apj, 803, 60,
  \dodoi{10.1088/0004-637X/803/2/60}

\bibitem[{{Saito} {et~al.}(2016){Saito}, {Iono}, {Xu}, {Ueda}, {Nakanishi},
  {Yun}, {Kaneko}, {Yamashita}, {Lee}, {Espada}, {Motohara}, \&
  {Kawabe}}]{Saito16}
{Saito}, T., {Iono}, D., {Xu}, C.~K., {et~al.} 2016, \pasj, 68, 20,
  \dodoi{10.1093/pasj/psv136}

\bibitem[{{Saito} {et~al.}(2017){Saito}, {Iono}, {Xu}, {Sliwa}, {Ueda},
  {Espada}, {Kaneko}, {K{\"o}nig}, {Nakanishi}, {Lee}, {Yun}, {Aalto},
  {Hibbard}, {Yamashita}, {Motohara}, \& {Kawabe}}]{Saito17b}
---. 2017, \apj, 835, 174, \dodoi{10.3847/1538-4357/835/2/174}

\bibitem[{{Saitoh} {et~al.}(2009){Saitoh}, {Daisaka}, {Kokubo}, {Makino},
  {Okamoto}, {Tomisaka}, {Wada}, \& {Yoshida}}]{Saitoh09}
{Saitoh}, T.~R., {Daisaka}, H., {Kokubo}, E., {et~al.} 2009, \pasj, 61, 481,
  \dodoi{10.1093/pasj/61.3.481}

\bibitem[{{Sanders} {et~al.}(1991){Sanders}, {Scoville}, \&
  {Soifer}}]{Sanders91}
{Sanders}, D.~B., {Scoville}, N.~Z., \& {Soifer}, B.~T. 1991, \apj, 370, 158,
  \dodoi{10.1086/169800}

\bibitem[{{Sandstrom} {et~al.}(2013){Sandstrom}, {Leroy}, {Walter}, {Bolatto},
  {Croxall}, {Draine}, {Wilson}, {Wolfire}, {Calzetti}, {Kennicutt}, {Aniano},
  {Donovan Meyer}, {Usero}, {Bigiel}, {Brinks}, {de Blok}, {Crocker}, {Dale},
  {Engelbracht}, {Galametz}, {Groves}, {Hunt}, {Koda}, {Kreckel}, {Linz},
  {Meidt}, {Pellegrini}, {Rix}, {Roussel}, {Schinnerer}, {Schruba}, {Schuster},
  {Skibba}, {van der Laan}, {Appleton}, {Armus}, {Brandl}, {Gordon}, {Hinz},
  {Krause}, {Montiel}, {Sauvage}, {Schmiedeke}, {Smith}, \&
  {Vigroux}}]{Sandstrom13}
{Sandstrom}, K.~M., {Leroy}, A.~K., {Walter}, F., {et~al.} 2013, \apj, 777, 5,
  \dodoi{10.1088/0004-637X/777/1/5}

\bibitem[{{Sch{\"o}ier} {et~al.}(2005){Sch{\"o}ier}, {van der Tak}, {van
  Dishoeck}, \& {Black}}]{Schoier05}
{Sch{\"o}ier}, F.~L., {van der Tak}, F.~F.~S., {van Dishoeck}, E.~F., \&
  {Black}, J.~H. 2005, \aap, 432, 369, \dodoi{10.1051/0004-6361:20041729}

\bibitem[{{Scoville} {et~al.}(2016){Scoville}, {Sheth}, {Aussel}, {Vanden
  Bout}, {Capak}, {Bongiorno}, {Casey}, {Murchikova}, {Koda},
  {{\'A}lvarez-M{\'a}rquez}, {Lee}, {Laigle}, {McCracken}, {Ilbert}, {Pope},
  {Sanders}, {Chu}, {Toft}, {Ivison}, \& {Manohar}}]{Scoville16}
{Scoville}, N., {Sheth}, K., {Aussel}, H., {et~al.} 2016, \apj, 820, 83,
  \dodoi{10.3847/0004-637X/820/2/83}

\bibitem[{{Sliwa} {et~al.}(2014){Sliwa}, {Wilson}, {Iono}, {Peck}, \&
  {Matsushita}}]{Sliwa14}
{Sliwa}, K., {Wilson}, C.~D., {Iono}, D., {Peck}, A., \& {Matsushita}, S. 2014,
  \apjl, 796, L15, \dodoi{10.1088/2041-8205/796/1/L15}

\bibitem[{{Sliwa} {et~al.}(2017){Sliwa}, {Wilson}, {Matsushita}, {Peck},
  {Petitpas}, {Saito}, \& {Yun}}]{Sliwa17}
{Sliwa}, K., {Wilson}, C.~D., {Matsushita}, S., {et~al.} 2017, \apj, 840, 8,
  \dodoi{10.3847/1538-4357/aa689b}

\bibitem[{{Solomon} \& {Vanden Bout}(2005)}]{Solomon05}
{Solomon}, P.~M., \& {Vanden Bout}, P.~A. 2005, \araa, 43, 677,
  \dodoi{10.1146/annurev.astro.43.051804.102221}

\bibitem[{{Teyssier} {et~al.}(2010){Teyssier}, {Chapon}, \&
  {Bournaud}}]{Teyssier10}
{Teyssier}, R., {Chapon}, D., \& {Bournaud}, F. 2010, \apjl, 720, L149,
  \dodoi{10.1088/2041-8205/720/2/L149}

\bibitem[{{Tomi{\v{c}}i{\'c}} {et~al.}(2018){Tomi{\v{c}}i{\'c}}, {Hughes},
  {Kreckel}, {Renaud}, {Pety}, {Schinnerer}, {Saito}, {Querejeta}, {Faesi}, \&
  {Garcia-Burillo}}]{Tomicic18}
{Tomi{\v{c}}i{\'c}}, N., {Hughes}, A., {Kreckel}, K., {et~al.} 2018, \apjl,
  869, L38, \dodoi{10.3847/2041-8213/aaf810}

\bibitem[{{U} {et~al.}(2012){U}, {Sanders}, {Mazzarella}, {Evans}, {Howell},
  {Surace}, {Armus}, {Iwasawa}, {Kim}, {Casey}, {Vavilkin}, {Dufault},
  {Larson}, {Barnes}, {Chan}, {Frayer}, {Haan}, {Inami}, {Ishida},
  {Kartaltepe}, {Melbourne}, \& {Petric}}]{U12}
{U}, V., {Sanders}, D.~B., {Mazzarella}, J.~M., {et~al.} 2012, \apjs, 203, 9,
  \dodoi{10.1088/0067-0049/203/1/9}

\bibitem[{{van der Tak} {et~al.}(2007){van der Tak}, {Black}, {Sch{\"o}ier},
  {Jansen}, \& {van Dishoeck}}]{vanderTak07}
{van der Tak}, F.~F.~S., {Black}, J.~H., {Sch{\"o}ier}, F.~L., {Jansen}, D.~J.,
  \& {van Dishoeck}, E.~F. 2007, \aap, 468, 627,
  \dodoi{10.1051/0004-6361:20066820}

\bibitem[{Virtanen {et~al.}(2020)Virtanen, Gommers, Oliphant, Haberland, Reddy,
  Cournapeau, Burovski, Peterson, Weckesser, Bright, {van der Walt}, Brett,
  Wilson, Millman, Mayorov, Nelson, Jones, Kern, Larson, Carey, İlhan Polat,
  Feng, Moore, VanderPlas, Laxalde, Perktold, Cimrman, Henriksen, Quintero,
  Harris, Archibald, Ribeiro, Pedregosa, {van Mulbregt}, Nunez-Iglesias, \&
  {SciPy 1.0 Contributors}}]{Virtanen20}
Virtanen, P., Gommers, R., Oliphant, T., {et~al.} 2020, Nature Methods, 17,
  261, \dodoi{10.1038/s41592-019-0686-2}

\bibitem[{{Watanabe} {et~al.}(2014){Watanabe}, {Sakai}, {Sorai}, \&
  {Yamamoto}}]{Watanabe14}
{Watanabe}, Y., {Sakai}, N., {Sorai}, K., \& {Yamamoto}, S. 2014, \apj, 788, 4,
  \dodoi{10.1088/0004-637X/788/1/4}

\bibitem[{{Whitmore} {et~al.}(1999){Whitmore}, {Zhang}, {Leitherer}, {Fall},
  {Schweizer}, \& {Miller}}]{Whitmore99}
{Whitmore}, B.~C., {Zhang}, Q., {Leitherer}, C., {et~al.} 1999, \aj, 118, 1551,
  \dodoi{10.1086/301041}

\end{thebibliography}
\bibliographystyle{aasjournal}

\begin{deluxetable*}{llcccccccccccc}
\tabletypesize{\scriptsize}
\tablecaption{Log of ALMA Observations \label{table_obs}}
\tablewidth{0pt}
\tablehead{
Band &UT Date &\multicolumn{3}{c}{Configuration} &$T_{\rm sys}$ &\multicolumn{2}{c}{Spectral Window} &$t_{\rm integ}$ &\multicolumn{3}{c}{Calibrator} \\
 \cline{3-5} \cline{7-8} \cline{10-12}
 & &FoV &$N_{\rm ant}$ &$L_{\rm baseline}$ & &USB &LSB & &Flux &Bandpass &Phase \\
 & &(\arcsec) & &(m) &(K) &(GHz) &(GHz) &(min.) & & & 
}
\startdata
B3 &2014 Jun. 28 &57.2 &29 &19--639 &40--\phantom{0}70 &107.529 &\phantom{0}95.669 &26.6 &Ceres &J1058+0133 &J1008+0621 \\
B3 &2015 Mar. 9 &54.7 &30 &15--328 &55--155 &112.487 &100.414 &28.1 &Ganymede &J1058+0133 &J1011-0423 \\
B3 &2015 May. 16 &54.7 &41 &21--558 &40--100 &112.480 &100.406 &23.0 &J1058+015 &J1058+0133 &J1011-0423 \\
B6 &2014 Dec. 8 &28.6 &35 &15--349 &65--115 &229.970 &215.805 &17.5$\times$3 &Ganymede &J1037-2934 &J1011-0423 \\
B6 &2014 Dec. 10 &27.3 &38 &15--349 &65--135 &242.443 &227.702 &14.7$\times$3 &Callisto &J1037-2934 &J1011-0423 
\enddata
%% Any table notes must follow the \end{tabular} command.
\tablecomments{
Column 3-5: FoV, $N_{\rm ant}$, and $L_{\rm baseline}$ are the FWHM of the ALMA primary beam, number of 12~m antennas, and projected baseline length of the assigned configuration, respectively.
Column 7-8: Central frequency of the two continuous spws in upper and lower sidebands (USB and LSB).
Column 9: Total integration time on NGC~3110.
The Band~6 observations have three pointing fields due to the small FoV.
Column 10-12: Assigned calibrators.
}
\end{deluxetable*}

\begin{deluxetable*}{lcccccccccccc}
\tabletypesize{\scriptsize}
\tablecaption{Line and Imaging Properties \label{table_line}}
\tablewidth{0pt}
\tablehead{
Line &$\nu_{\rm obs}$ &$E_{\rm u}$/k &$uv$-weight &Beam Size (P.A.) &$V_{\rm ch}$ &$\sigma_{\rm ch}$ &$S_{\rm line}\Delta v$ \\
 &(GHz) &(K) & &(\arcsec (\degr)) &(km s$^{-1}$) &(mJy beam$^{-1}$) &(Jy km s$^{-1}$)
}
\startdata
\co~($J$ = 1--0) &113.360 &\phantom{0}5.5 &natural &1.81 $\times$ 1.43 ($-$82) &20 &1.2 &303.34 $\pm$ 1.91 \\
\co~($J$ = 2--1) &226.716 &16.6 &natural &1.68 $\times$ 0.93 (+74) &20 &1.2 &854.84 $\pm$ 1.83 \\
\tco~($J$ = 1--0) &108.374 &\phantom{0}5.3 &natural &1.16 $\times$ 1.14 ($-$32) &20 &1.1 &\phantom{0}10.08 $\pm$ 0.69 \\
\tco~($J$ = 2--1) &216.744 &15.9 &natural &1.59 $\times$ 0.93 (+80) &20 &0.7 &\phantom{0}38.01 $\pm$ 0.57 \\
\ceo~($J$ = 2--1) &215.920 &15.8 &natural &1.59 $\times$ 0.93 (+80) &40 &0.4 &\phantom{00}7.59 $\pm$ 0.25
%\co~($J$ = 1--0) &113.360 &\phantom{0}5.5 &natural &1.81 $\times$ 1.43 ($-$82) &20 &1.2 &258.41 $\pm$ 2.21 \\
%\co~($J$ = 2--1) &226.716 &16.6 &natural &1.68 $\times$ 0.93 (+74) &20 &1.2 &723.97 $\pm$ 1.77 \\
%\tco~($J$ = 1--0) &108.374 &\phantom{0}5.3 &natural &1.16 $\times$ 1.14 ($-$32) &20 &1.1 &\phantom{00}4.10 $\pm$ 0.45 \\
%\tco~($J$ = 2--1) &216.744 &15.9 &natural &1.59 $\times$ 0.93 (+80) &20 &0.7 &\phantom{0}31.46 $\pm$ 0.79 \\
%\ceo~($J$ = 2--1) &215.920 &15.8 &natural &1.59 $\times$ 0.93 (+80) &40 &0.4 &\phantom{00}5.90 $\pm$ 0.33
\enddata
%% Any table notes must follow the \end{tabular} command.
\tablecomments{
Column 6: Velocity resolution of the data cube.
Column 7: Noise rms per channel per pixel in the data which have velocity resolution of $\Delta v$.
Column 8: Total integrated intensity. All the measurements are done after convolving the synthesized beam to 2\farcs0.
We only consider the statistical error here.
}
\end{deluxetable*}

\appendix

\section{Scatter plots with varying \texorpdfstring{\aCO}{aCO}}\label{App1}
\begin{figure*}
\begin{center}
\includegraphics[width=15cm]{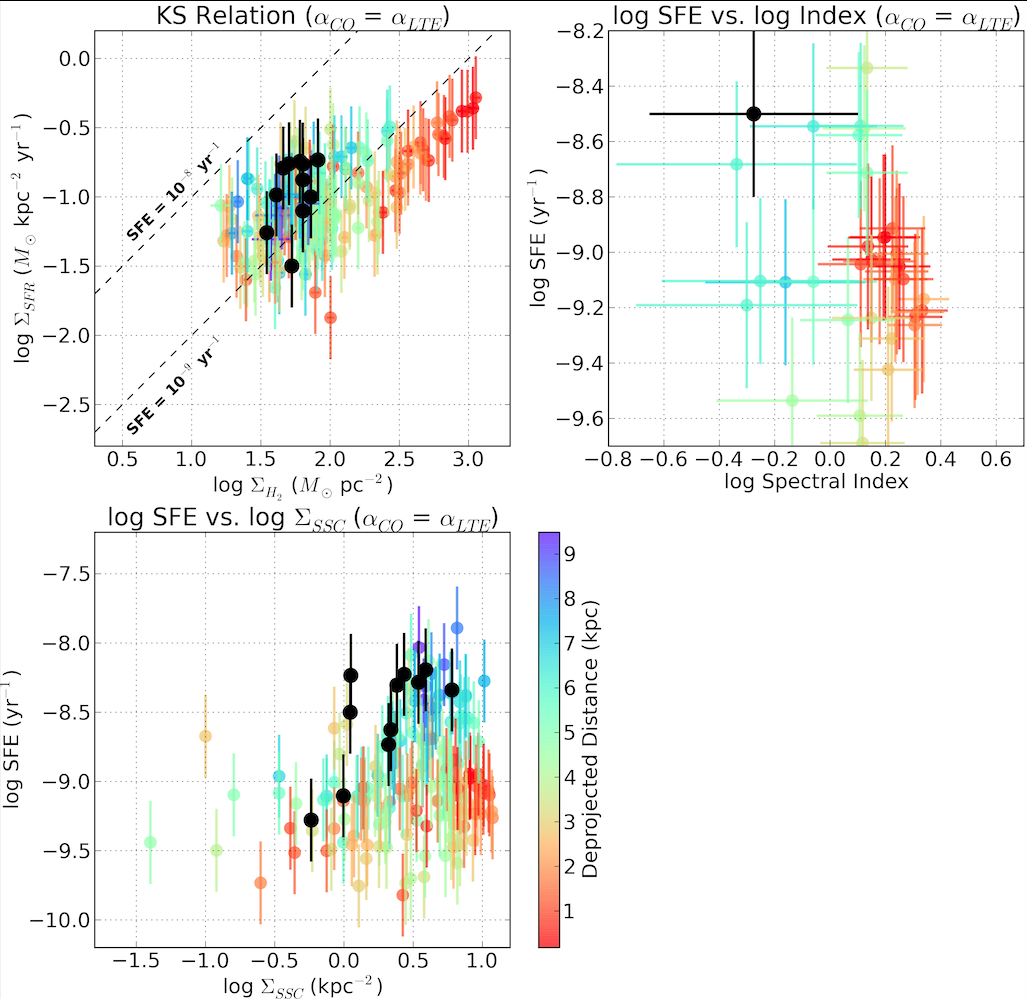}
\end{center}
\caption{Same as the plots in Figure~\ref{fig_sfe}, but here we apply \aLTE($T_{\rm rot}$) instead of a constant \aCO\ of 1.7~\uaco.
}\label{fig_appendix}
\end{figure*}

We show the scatter plots including the KS relation, SFE vs. spectral index, and SFE vs SSC density in Figure~\ref{fig_sfe}. The molecular gas masses are derived assuming a constant \aCO\ of 1.7~\uaco\ (see Section~\ref{alpha}). In order to discuss how these plots change when we apply the spatially varying \aCO, we reconstruct these with \aLTE($T_{\rm rot}$) in Figure~\ref{fig_appendix}. The number of data points decrease compared to Figure~\ref{fig_sfe}, as the \aLTE($T_{\rm rot}$)\ estimate is limited for \tcotwo-detected apertures. However, we confirm that the varying \aCO\ does not strongly affect the trends of the region ``A" discussed in the main text. Thus, both the constant and varying \aCO\ measurements lead to the same conclusions of this paper.

\listofchanges

\end{document}